\documentclass{jpp}
\usepackage{graphicx}

\usepackage[utf8]{inputenc}
\usepackage[T1]{fontenc}
\usepackage{amsmath}
\usepackage{orcidlink}
\usepackage{hyperref}
\usepackage{subcaption}

\shorttitle{Collision operator for electron runaway in cold weakly-ionized plasmas}
\shortauthor{Y. Lee, P. Aleynikov, P.C. de Vries, J.-K. Park and Y.-S. Na}

\title{Collision operator for electron runaway in cold weakly-ionized plasmas}

\author{Yeongsun Lee\aff{1,2}\orcidlink{0000-0003-4474-416X},
 Pavel Aleynikov\aff{3}\orcidlink{0009-0002-3037-3679},
 Peter de Vries\aff{4}\orcidlink{0000-0001-7304-5486},
 Jong-Kyu Park\aff{1,5}\orcidlink{0000-0003-2419-8667}
 \and Yong-Su Na\aff{1}\orcidlink{0000-0001-7270-3846}
 \corresp{\email{ysna@snu.ac.kr}}}

\affiliation{\aff{1}Department of Nuclear Engineering, Seoul National University, Seoul, South Korea,
\aff{2}Nuclear Research Institute for Future Technology and Policy, Seoul National University, Seoul, South Korea,
\aff{3}Max-Planck Institute fur Plasmaphysik, Greifswald, Germany,
\aff{4}ITER Organization, Route de Vinon sur Verdon, CS 90 046, 13067 St Paul Lez Durance, France,
\aff{5}Princeton Plasma Physics Laboratory, Princeton, New Jersey 08543, USA}

\begin{document}

\maketitle

\begin{abstract}
%Dreicer generation is one of main mechanisms of runaway electrons generation. 
In cold weakly-ionized plasmas, Dreicer generation mechanism can be non-diffusive as demonstrated in [Y. Lee et. al. \textit{Phys. Rev. Lett.} 133 \textbf{17} 175102 (2024)]. By expanding the previous letter, we present the detailed description of a proper collision operator to precisely account for the non-diffusive electron kinetics. The operator appropriately combines the Fokker-Planck operator and Boltzmann operator where free-bound collision cross sections are valid in low energy region. 
The proposed operator is envisaged to predict runaway electrons generations in cold weakly-ionized plasmas, particularly to design a runaway-free reactor tokamak startup.
\end{abstract}

\section{Introduction}
In weakly-ionized plasmas, inelastic collisions play key roles in Runaway Electron (RE) \citep{Wilson1925MPCPS} generation by slowing electrons down \citep{Gurevich1961JETP} and creating knock-on sources \citep{Rosenbluth1997NF}. For highly energetic electrons, inelastic slowing down can be adequately described by the Fokker-Planck (FP) operator \citep{Rosenbluth1957PR, Chandrasekhar1943RMP} since the typical energy transfer is much smaller than their energy \citep{Breizman2019NF}. Similarly, free-bound knock-on collisions, often approximated as free-free knock-on under the assumption of negligible ionization potential \citep{McDevitt2019PPCF}, can be accounted for by the Boltzmann operator \citep{Rosenbluth1997NF, Landau1981}. Main interest of previous studies on free-bound collision operators \citep{Kirillov1975, Mosher1975PoF, Hesslow2017PRL, Hesslow2018JPP, Breizman2019NF} has concentrated on dynamics of energetic electrons to comprehend undesirable disruption REs \citep{Lehnen2008PRL, Martin-solis2017NF, Reux2021PRL} due to the danger of triggering plasma instabilities \cite{Snipes2008NF, Liu2023PRL} and ability to damage the device \citep{Nygren1997JNM, Matthews2016PS, Vries2023NF}.

Startup REs \citep{Knoepfel1975PRL, Knoepfel1979NF,Sharma1988NF} are recently well highlighted for upcoming ITER plasma initiation \citep{Vries2019NF, Vries2023NF} since significant current carried by them complicates the startup or even leads to failure \citep{Gribov2018EPS, Vries2020PPCF, Hoppe2022JPP, Matsuyama2022NF, Lee2023NF, Lee2024PRL, deVries2025NF}. Yet, the collision models suitable for describing disruption REs may be inadequate for analyzing them. For instance, energy of startup REs is insufficient to make use of the 1st order Born-approximation \citep{Landau2013} essential for the Bethe stopping power \citep{Bethe1930AP} and Thomas-Fermi (TF) models \citep{Kirillov1975, Hesslow2017PRL}. That is, valid collision cross sections are necessary for both elastic and inelastic collisions. In addition, energy transfer via inelastic collisions are predominantly hard when incident electron energy is close to the ionization potential. This necessitates individual inelastic collisions to be accounted by the binary Boltzmann operator \cite{Landau2013}. Furthermore, the background Maxwellian assumption needs to be verified under a strong effect of neutrals on electrons in order to adopt the linearized free-free collision operator \citep{Helander2005Cam}.

According to the classical Dreicer kinetics \citep{Gurevich1961JETP, Connor1975NF}, the key phase space region determining the formation of Dreicer flow is the narrow singular layer across the critical momentum where the energy diffusion drives an upward flow. However, if the critical energy is close to the ionization potential in cold weakly-ionized plasmas, the key region can be wider from the minimum excitation energy to critical energy in which hard inelastic collisions should be considered by the binary Boltzmann operator. Indeed, it was recently elucidated that the binary nature of inelastic collisions allows quasi-frictionless acceleration for some particles and facilitates Dreicer generation \citep{Dreicer1959PR}, yielding significant enhancement in the generation rate \citep{Lee2024PRL}. %, renders its necessity clearer.
The collision operator aided by the correct cross sections and Boltzmann operator was originally presented in \citep{Lee2024PRL} to include this nature in designing a runaway-free reactor startup scenario with better confidence, where plasma parameters remain far away from the RE generation parametric region. This paper expands on the collision operator in \citep{Lee2024PRL} by (1) describing the model development for electron-hydrogen atom collisions in greater detail, (2) explicitly verifying the background Maxwellian assumption in cold weakly-ionized plasmas under the consideration of complete linearized Coulomb operator and (3) presenting the numerical method for implementation of this model into kinetic solvers using Finite Volume Method (FVM).

The paper is organized as follows. In section \ref{sec2}, we present the Fokker-Planck-Boltzmann (FPB) operator with appropriate free-bound collision cross sections. In section \ref{sec3}, the numerical implementation of this operator will be clarified. In section \ref{sec4}, the operator is applied to solve Dreicer problem.
 
\section{Kinetic model \label{sec2}}
The kinetic equation for the electrons is
\begin{equation}
    \frac{\partial f_e}{\partial t} + eE\Big[ \frac{\mu}{p^2} \frac{\partial}{\partial p} p^2 + \frac{1}{p} \frac{\partial}{\partial \mu} (1-\mu^2) \Big] f_e = \mathcal{C} \{ f_e \} \label{eq:kin_eq}
\end{equation}
where $f_e(p,\mu)$ is the electron distribution function, $p$ is particle momentum and $\mu=\cos{\theta}$ is cosine of pitch-angle $\theta$. Suppose the pure weakly-ionized plasmas and assume hydrogen is in the ground state. There are two types of interactions depending on colliding species. Charged particles interact each other through the long-range Coulomb interaction. The FP operator appropriately describes such inverse-squared forces \citep{Rosenbluth1957PR}; the free-free knock on collisions \citep{Rosenbluth1957PR, Chiu1998NF} are negligible due to low ionization fraction. However, the interaction between electron and hydrogen atom is short-range (close) due to the screened charge. We apply the FPB operator to account for electron-hydrogen atom interactions,
\begin{equation}
    \mathcal{C} \{ f_e \} = \mathcal{C}^{eH} \{ f_e \} + \mathcal{C}_{FP} \{ f_e \}
\end{equation}
where $\mathcal{C}^{eH} \{ f_e \}$ is the FPB operator for electron-H atom interactions and $\mathcal{C}_{FP} \{ f_e \}$ is the FP operator for collisions with charged particles. The FPB operator consists of two parts
\begin{equation}
    \mathcal{C}^{eH} \{ f_e \} = \mathcal{C}^{soft}_{FP} \{ f_e \} + \mathcal{B}^{hard} \{ f_e \} \label{eq:full_kin}
\end{equation}
where $\mathcal{C}^{soft}_{FP} \{ f_e \}$ is the FP operator which includes all elastic collisions and soft inelastic collisions and $\mathcal{B}^{hard} \{ f_e \}=\mathcal{B}^{hard}_{iz} \{ f_e \} + \mathcal{B}^{hard}_{ex} \{ f_e \}$ is the Boltzmann operator only accounting for hard inelastic collisions. The subscripts $iz$ and $ex$ denote ionizing and exciting collisions, respectively. In $\mathcal{C}^{soft}_{FP} \{ f_e \}$, the integral boundary of Coulomb logarithm is appropriately selected to avoid the double counting of collisions \citep{Embreus2018JPP}. 

A collision is soft if the accompanied energy loss is small compared to the incident electron energy and otherwise hard. We introduce the soft-hard separation factor $h$ for the systematic categorization. If the energy loss ratio of the incident electron is higher than $h$, such collisions are hard and included in $\mathcal{B}^{hard} \{ f_e \}$; otherwise, they are described by $\mathcal{C}^{soft}_{FP} \{ f_e \}$. 

Let $T$ is the kinetic energy of the incident electron, $W$ is that of the ejected electron, $T-W$ is that of the scattered electron and lowercase representation ($t$, $w$ and $t-w$) indicates normalization by the binding energy of the ejected electron $B$. We adopt the convention to call the fast one \textit{scattered} and the slower one \textit{ejected} after ionizing collisions due to indistinguishability of electrons, i.e. interchanging interpretation of $W$ and $T-W$ if $W>T-W$. The energy loss ratio of the incident electron is $\min\{(1+w), (t-w)\}/t$ under this convention,
\begin{eqnarray}
    \text{Collision \ is \ hard if} &\ & \frac{\min\{(1+w), (t-w)\}}{t} > h \notag \\
    \text{Collision \ is \ soft if} &\ & \frac{\min\{(1+w), (t-w)\}}{t} < h \notag.
\end{eqnarray}
The scattered and ejected electron is regarded as the test and field particle in kinetic treatment.

This section is organized as follows. $\mathcal{B}^{hard}$ is derived in Sec. \ref{sec2-1} and $\mathcal{C}^{soft}_{FP}$ in Sec. \ref{sec2-2}. In Sec. \ref{sec2-3}, invariance in $h$ is proven. Sec. \ref{sec2-4} is about $\mathcal{C}_{FP}$. Electron acceleration mechanism is discussed in Sec. \ref{sec2-5}.

\subsection{Boltzmann collision model \label{sec2-1}}
\subsubsection{Hard ionizing collisions}
Inelastic collisions accounted by the Botlzmann operator require the \textit{differential} cross section, which informs the distribution of the ejected electrons. We apply RBED model \cite{Kim2000PRA} that is the relativistic extension of the BED model \cite{Kim1994PRA}. Note that the BED model successfully reproduced the experimentally measured total ionization cross section \cite{Kim1994PRA}. In addition, energy distribution of secondary (ejected) electrons of the BED model has a good agreement with that of experimental data \cite{Kim1994PRA}. Therefore, this model is appropriate to use in our purpose.

Singly Differential Cross Section (SDCS) of RBED model is
\begin{align}
    &\frac{\partial \sigma_{iz}(W, T)}{\partial W} = \frac{4\pi a_0^2 \alpha^4 N}{(\beta_t^2 + \beta_u^2 + \beta_b^2)2b'} \notag \\
    &\times\{ \frac{(N_i/N)-2}{t+1}(\frac{1}{w+1} + \frac{1}{t-w}) \frac{1+2t'}{(1+t'/2)^2} \notag \\
    &+[2-(N_i/N)][\frac{1}{(w+1)^2} + \frac{1}{(t-w)^2} + \frac{b'^2}{(1+t'/2)^2}] \notag \\
    &+ \frac{1}{N(w+1)} \frac{df(w)}{dw} [\ln (\frac{\beta_t^2}{1-\beta_t^2}) -\beta_t^2 - \ln (2b')] \} \label{eq:SDCS}
\end{align}
where we borrow the notation of Ref \citep{Kim2000PRA}. This model is developed by combining the Mott cross section \citep{Mott1930} and Bethe cross section \citep{Bethe1930AP}. The Bethe cross section accounts for the dipole contribution, the last line in Eqn. (\ref{eq:SDCS}) proportional to $\frac{1}{w+1}\frac{df(w)}{dw}$, which is unfortunately not symmetric under interchanging two electrons after the collision. %Given that the inelastic collision is intrinsically symmetric in the interchange due to electron 
Due to indistinguishability of electrons, we only use an ejected electron part of Eqn. (\ref{eq:SDCS}) including description of a scattered electron part, i.e. $(\partial \sigma_{iz}) / (\partial W) (W,T) = (\partial \sigma_{iz}) / (\partial W) (T-W-B,T) \ \text{if } W > (T-B)/2$.

The SCDS doesn't contain the scattering angle distribution $f_{\tilde{\mu}}(\vec{p}_1, \vec{p})$. When electron energy is low, the pitch-angle scattering is strong. In addition, the free-bound hard collisions by energetic electrons can be approximated as the free-free knock-on collisions \citep{Moller1932AP, Breizman2019NF}. Therefore, we prescribe $f_{\tilde{\mu}}(\vec{p}_1, \vec{p})$ by interpolating the isotropic and Moller scattering angle distribution \citep{Moller1932AP},
\begin{eqnarray}
    f_{\tilde{\mu}}(\vec{p}_1, \vec{p}) = \frac{\frac{1}{2} + (\delta(\tilde{\mu}-\mu^*)-\frac{1}{2})sig(\ln (p/0.1mc))}{2\pi} \label{eq:scat_ang}
\end{eqnarray}
where $\mu^*=\sqrt{\frac{(\gamma_1+1)(\gamma-1)}{(\gamma_1-1)(\gamma+1)}}$ is cosine of the Moller scattering angle, $\tilde{\mu}=(\vec{p}_1 \cdot \vec{p}) / (p_1 p)$ is cosine of deflection angle and $\delta$ is the delta function. The sigmoid function transits $f_{\tilde{\mu}}(\vec{p}_1, \vec{p})$ to the relativistic limit ($\delta(\tilde{\mu}-\mu^*)/2\pi$) across $p=0.1$. Note Eqn. (\ref{eq:scat_ang}) satisfies $\int d\mu d \phi f_{\tilde{\mu}}(\vec{p}_1, \vec{p}) = 1$.

The number of hard ionizing collisions by incident electrons with $\vec{p}_1$ that produces electrons with $\vec{p}$ in phase space volume $d^3\vec{p}_1 d^3 \vec{p}$ during $dt$ is
\begin{align}
    (dn)^{hard}_{iz} d \vec{p}_1 d \vec{p} &= \mathcal{H} (p_1 - p_{bnd}) f_e(\vec{p}_1) d \vec{p}_1 [\frac{\partial \sigma (T, W)}{\partial W} f_{\tilde{\mu}}(\vec{p}_1, \vec{p}) dW d\mu d\phi] v_1 n_H dt \label{eq:dn_iz}
\end{align}
where $n_H$ is hydrogen density. The produced electrons include the test particles that are reborn after losing the energy through the hard ionizing collisions and the field particles created by all ionizing collisions. When $t > 2w + 1$ is met, the produced electrons are the field particles. The associated momentum $mc\sqrt{(2w'+b'+1)^2-1}$ is the minimum of $p_1$. In addition, there is no ionizing collision if $t < w + 1$. In this case, the corresponding momentum $mc\sqrt{(w'+b'+1)^2-1}$ is the maximum of $p_1$. Furthermore, the hard collision condition demands $t > \frac{w}{1-h}$, i.e. $p_1 > mc\sqrt{(\frac{w'}{1-h}+1)^2-1}$. The three constraints for $p_1$ yield the lower boundary for hard ionizing collisions,
\begin{align}
    p_{bnd} &= mc \Big\{ \min_2\Big(\sqrt{(w'+b'+1)^2-1}, \sqrt{(\frac{w'}{1-h}+1)^2-1}, \sqrt{(2w'+b'+1)^2-1}\Big) \Big\}
\end{align}
where
\begin{equation}
    \min_2(a,b,c) \equiv \min(a, \max(b,c))
\end{equation}
is the second minimum function.

Boltzmann operator can be obtained by taking an integral $\int d \vec{p}_1$ and dividing by $dt d \vec{p}$
\begin{align}
    \mathcal{B}^{hard}_{iz} \{f_e\} &= \frac{\int_{\vec{p}_1} \Big[ (dn)^{hard}_{iz} d \vec{p}_1 d \vec{p} \Big]}{dt d\vec{p}} - \frac{\int_{\vec{p}_1} \Big[(dn)^{hard}_{iz} d \vec{p}_1 d \vec{p} \Big] (\vec{p} \leftrightarrow \vec{p}_1)}{dt d\vec{p}}
\end{align}
where $(\vec{p} \leftrightarrow \vec{p}_1)$ is the interchange operator of $\vec{p}$ and $\vec{p}_1$. This can be rewritten as
\begin{align}
    \mathcal{B}^{hard}_{iz} \{f_e\} &= n_H \frac{v}{p^2} \int_{p_1 \ge p_{pnd}} d\vec{p}_1 \Big[f_e(\vec{p}_1) v_1 \frac{\partial \sigma_{iz} (W,T)}{\partial W} f_{\tilde{\mu}} (\vec{p}_1, \vec{p}) \Big] \notag - \nu_{iz}^{hard} f_e (\vec{p})
\end{align}
where, $dW/dp = v$ is used, $\nu_{iz}^{hard} = n_H v \sigma_{iz}^{hard}(T)$ is the hard ionizing collision frequency with the total cross section of hard ionizing collisions $\sigma_{iz}^{hard}(T)$
\begin{equation}
    \sigma_{iz}^{hard}(T) = B\int_{w_{bnd}}^{\frac{t-1}{2}} dw \frac{\partial \sigma_{iz}(W,T)}{\partial W}.
\end{equation}
Estimating the total cross section has the upper boundary given by $(t-1)/2$ because collisions are doubly counted above this. Moreover, there is no ionizing collision for $w<0$. The hard collision condition yields $w > ht-1$ provided that $w$ is of the ejected particle between $0$ and $(t-1)/2$. The resulting integral boundary $w_{bnd}$ is
\begin{equation}
    w_{bnd} = \min_2 \Big(0, ht-1, \frac{t-1}{2}\Big).
\end{equation}

After multiplying $p^2$, take an integral $\int d\phi$, and represent $\mathcal{B}_{iz}$ using $F_e (p, \mu) = \int d\phi p^2 f_e(\vec{p})$,
\begin{align}
    2\pi p^2 &\mathcal{B}^{hard}_{iz} \{f_e\} = n_H v \int_{p_{bnd}} dp_1 \int_{-1}^{1} d\mu_1 \Big[ F_e (p_1, \mu_1) v_1 \frac{\partial \sigma_{RBED}(W, T)}{\partial W} \Big] \notag \\
    &\times\Big(\frac{1}{2} + (\frac{\mathcal{H}((1-\mu^2)(1-\mu_1^2)-(\mu^*-\mu\mu_1)^2)}{\pi\sqrt{(1-\mu^2)(1-\mu_1^2)-(\mu^*-\mu\mu_1)^2}}- \frac{1}{2})sig (\ln (p/0.1mc))) \Big)  \notag \\
    &- \nu_{iz}^{hard} F_e (p, \mu).
\end{align}
with the help of the gyro-phase averaged Moller scattering angle \cite{Breizman2019NF}
\begin{equation}
    \int d\phi \delta (\tilde{\mu} - \mu^*) = \frac{2\mathcal{H}((1-\mu^2)(1-\mu_1^2)-(\mu^*-\mu\mu_1)^2)}{\sqrt{(1-\mu^2)(1-\mu_1^2)-(\mu^*-\mu\mu_1)^2}}.
\end{equation}

\subsubsection{Hard exciting collisions}
Hard exciting collisions are considered by the asymptotic expression of excitation cross section \citep{Stone2002JNIST}. This follows the BE-scaled Plane Wave Born (PWB) cross sections \citep{Kim2001PRA}, which has a good agreement with the CCC model \citep{Kim2001PRA}. The non-relativistic expression valid nearly upto $10 \ keV$ is appropripate for computing the runaway creation rate in the weakly-ionized plasmas due to the low critical energy for runaway condition. The BE-scaled PWB cross section do not include resonance effect \citep{Stone2002JNIST}. However, the error only exists in the limited region of phase space, i.e. at the near-threshold energy, and thus doesn't influences the electron acceleration significantly. The asymptotic cross section of electron impact excitation accompanying with $1s \to np$ transition has a form
\begin{equation}
    \sigma_{asympt}^{1s \to np}(T) = \frac{4\pi a_0^2 R}{T+B+E_{1n}} [a_{1n}\ln (T/R) + b_{1n} + c_{1n} R/T]
\end{equation}
where $a_{1n}, b_{1n}$ and $c_{1n}$ are coefficients of the asymptotic expression and $E_{1n}$ is the excitation energy for $1s \to np$ transition. The symbol of $\sigma_{asympt}^{1s \to np}(T)$ is replaced as $\sigma_{ex}^{1n}(T)$ below for brevity. Because the energy space region of incident electron to participate in hard exciting collisions is limited as $E_{1n} \leq T \leq E_{1n}/h$ due to the fixed excitation energy, the scattering angle distribution is presumed as isotropic. The number of hard exciting collisions involving with electron momentum transition from $\vec{p}_1$ to $\vec{p}$ in phase space volume $d\vec{p}_1 d\vec{p}$ during $dt$ is
\begin{align}
    (dn)^{hard}_{ex, 1s \to np} &d \vec{p}_1 d \vec{p} = f_e(\vec{p}_1) d \vec{p}_1 [\frac{\sigma_{ex}^{1n} (T)}{4\pi} \delta (T-W-E_{1n}) dW d\mu d\phi] v_1 n_H dt
\end{align}
Boltzmann operator can be obtained by
\begin{align}
    \mathcal{B}^{hard}_{ex, 1s\to np} \{f_e\} &= \frac{\int_{\vec{p}_1} \Big[ (dn)^{hard}_{ex, 1s \to np} d \vec{p}_1 d \vec{p} \Big]}{dt d\vec{p}} - \frac{\int_{\vec{p}_1} \Big[(dn)^{hard}_{ex, 1s \to np} d \vec{p}_1 d \vec{p} \Big] (\vec{p} \leftrightarrow \vec{p}_1)}{dt d\vec{p}}
\end{align}
This can be rewritten as
\begin{align}
    \mathcal{B}^{hard}_{ex, 1s\to np} \{f_e\} &= \frac{\nu_{ex, 1s\to np}^{hard}(p^+) F_e^0(p^+) }{4\pi p^2} \frac{v(p)}{v(p^+)}- \nu_{ex, 1s\to np}^{hard}(p) f_e(\vec{p})
\end{align}
where $p^+ = mc \sqrt{(T'+E_{1n}'+1)^2-1}$, $\nu_{ex, 1s\to np}^{hard}$ is the hard exciting collision frequency due to $1s$ to $np$ transition ($\nu_{ex, 1s\to np}^{hard}(p^+_{1n}) = n_H v(p^+_{1n}) \sigma_{ex}^{1n}(T+E_{1n})\mathcal{H} (\frac{E_{1n}}{T+E_{1n}} - h)$, $\nu_{ex, 1s\to np}^{hard}(p) = n_H v(p) \sigma_{ex}^{1n}(T)\mathcal{H} (\frac{E_{1n}}{T} - h)$).

Multiplying $p^2$, take an integral $\int d\phi$, and represent $\mathcal{B}^{hard}_{ex,1s\to np} \{ f_e \}=\mathcal{B}^{hard,+}_{ex,1s\to np} \{ f_e \}-\mathcal{B}^{hard,-}_{ex,1s\to np} \{ f_e \}$ :
\begin{align}
    2\pi p^2 \mathcal{B}^{hard}_{ex,1s\to np} \{ f_e \} &= \frac{\nu_{ex, 1s\to np}^{hard}(p^+_{1n}) }{2} \frac{v(p)}{v(p^+_{1n})} F_e^0(p^+_{1n})- \nu_{ex, 1s\to np}^{hard}(p) F_e(p,\mu).
\end{align}

We consider hard exciting collisions upto $1s \to 10p$ transition, i.e.
\begin{align}
    2\pi p^2 \mathcal{B}^{hard}_{ex} \{ f_e \} = \sum_{n=2}^{10} 2\pi p^2 \mathcal{B}^{hard}_{ex,1s\to np} \{ f_e \}.
\end{align}

\subsubsection{Normalization}
Let $\tilde{p} = p/mc$, $\tilde{t} = t/\tau_c$, $\tilde{\nu}_{iz}^{hard} = \nu_{iz}^{hard} \tau_c$ and $\tilde{\nu}_{ex}^{hard} = \nu_{ex}^{hard} \tau_c$, where $\tau_c \equiv \frac{4 \pi \varepsilon_0^2 m_e^2 c^3}{e^4 n_e ln \Lambda}$. Neglecting tilde, $2\pi p^2 \mathcal{B}^{hard} \{ f_e \}$ becomes
\begin{align}
    &2\pi p^2 \mathcal{B}^{hard} \{ f_e \} = n_H mc^3 \tau_c \frac{p}{\gamma} \Big[ \int_{p_{bnd}} dp_1 d\mu_1 \frac{d\sigma_{iz}}{dW} \notag \\
    &\times \Big( (\frac{\mathcal{H}((1-\mu^2)(1-\mu_1^2)-(\mu^*-\mu\mu_1)^2)}{\pi\sqrt{(1-\mu^2)(1-\mu_1^2)-(\mu^*-\mu\mu_1)^2}} - \frac{1}{2}) \frac{1}{1+\frac{0.1}{p}}+\frac{1}{2} \Big)  \frac{p_1}{\gamma_1}F_e(p_1, \mu_1) \Big] - \nu_{iz}^{hard} F_e(p, \mu) \notag \\
    &+ \sum_{n=2}^{10}\Big[ \frac{\nu_{ex, 1s \to np}^{hard}(p^+_{1n})}{2} \frac{\beta(p)}{\beta(p^+_{1n})} F_e^0(p^+_{1n}) - \nu_{ex, 1s \to np}^{hard}(p)  F_e(p, \mu) \Big] \label{eq:hard}.
\end{align}

\subsection{Fokker-Planck collisional model\label{sec2-2}}
\subsubsection{Elastic collisions}
\begin{figure}
    \centering
    \includegraphics[width=0.8\linewidth]{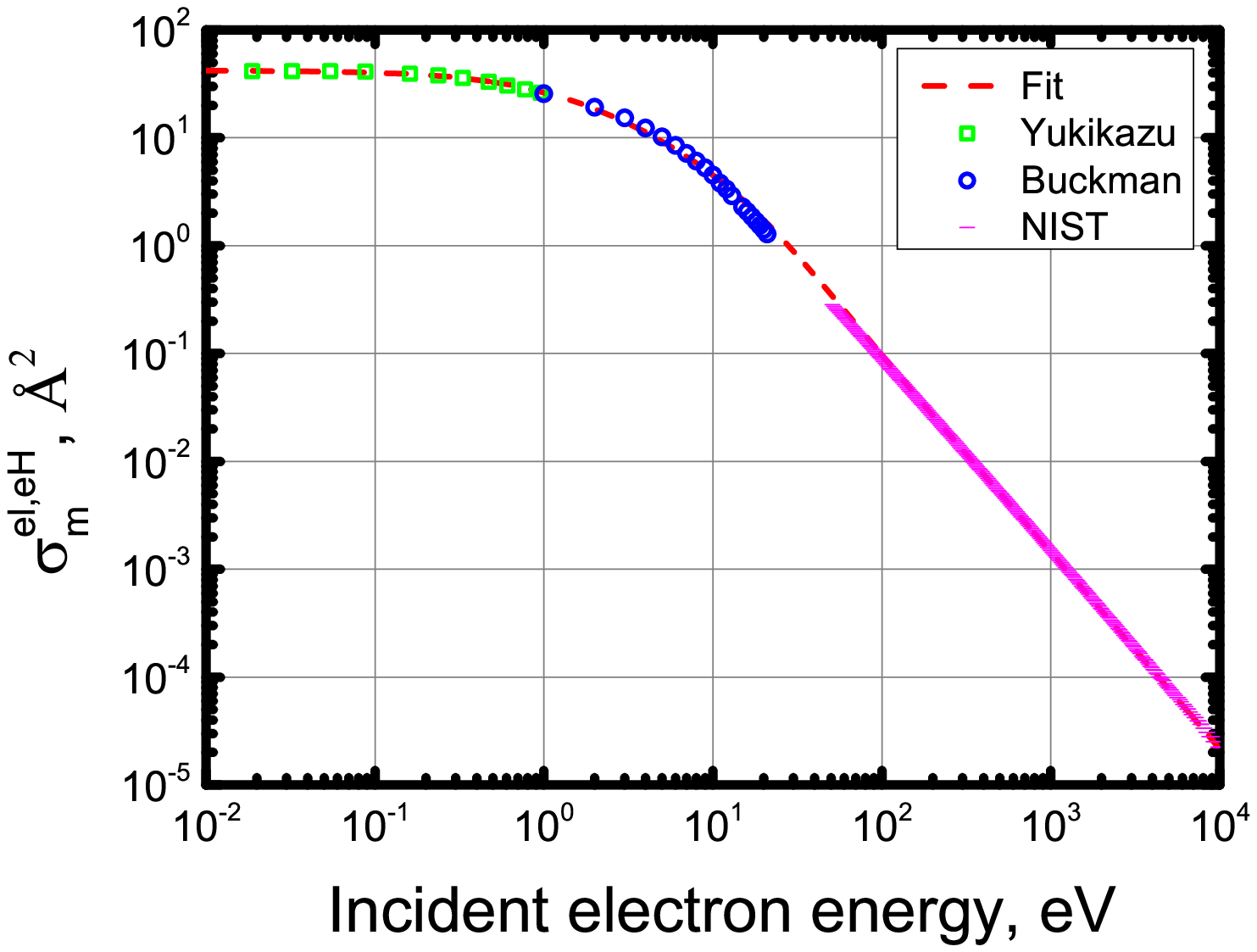}
    \caption{The momentum transfer cross section from the refined model with $\chi$ (red dashed curve), Ref. \citep{Itikawa1974ADNDT} (green square marker), Ref. \citep{Buckman2000ECA} (blue circle marker) and NIST standard database \citep{Jablonski2004JPCRD, Salvat2005CPC} (magenta bar).}
    \label{fig1}
\end{figure}
To account for elastic scattering of low energy electrons due to collision with hydrogen neutrals, we borrow a function form from the TF model \citep{Breizman2019NF} and refine it by fitting against experimental data \citep{Itikawa1974ADNDT, Buckman2000ECA, Jablonski2004JPCRD, Salvat2005CPC}. The direct usage of the TF model is not rigor because the TF model is valid for the partial screening effect of high Z partially-stripped ions/neutrals on energetic electrons: the majority of bound electrons need to have large principal quantum numbers and an incident electron should be energetic due to the 1st order Born-approximation \citep{Landau2013, Hesslow2018JPP}.

The function form of the elastic collision frequency is known as \citep{Breizman2019NF}
\begin{eqnarray}
    \nu_{el} = n_H c \frac{e^4}{4\pi \varepsilon_0^2 m^2 c^4} I_2 (y) \label{eq:freq_el}
\end{eqnarray}
where $I_2 (y)$ is the second screening coefficient with $y=274p/mc$. This coefficient grows logarithmically with $y$ at the high energy region, i.e. $I_2(y) \approx I_2(y_*) + \ln (y/y_*)$ with $y_*=26$. We extrapolate this to the low energy region, 
\begin{equation}
    I_2(y) \approx I_2(y_*) + \frac{\ln \Big[(\frac{y}{y_*})^4 + \exp(-I_2(y_*))^4\Big]}{4} \label{eq:I_2}.
\end{equation}
Let $\nu_{el}^{ref} \equiv \chi \nu_{el}$ be the refined frequency with a correction factor $\chi$. The outcome of refinement is displayed in Fig. \ref{fig1} by fitting the momentum transfer (transport) cross section $\sigma_m^{el,eH} \approx \chi \frac{e^4}{4\pi \varepsilon_0^2 m_e^2 v^4} I_2$ to those of experimental data \citep{Itikawa1974ADNDT, Buckman2000ECA} and the NIST Standard Database \citep{Jablonski2004JPCRD, Salvat2005CPC}. The red curve goes to the magenta bars for high energy particles while aligning well with experimental data at low energy range. In fact, $R^2\text{ score}$ is about $0.999$. The fitting finds,
\begin{equation}
     \chi= 1 + 
     \frac{a^{el} (511000 (\gamma-1))^{b^{el}}}{1+c^{el} (511000 (\gamma-1))^{b^{el}}} \label{eq:chi}
\end{equation}
where $a^{el}=79.837201$, $b^{el}=-1.0992754$ and $c^{el}=1.6387662$.

\subsubsection{Soft ionizing collisions}
Let the stopping power of soft ionizing collisions on electrons have the logarithmic factor
\begin{eqnarray}
    \Big( \frac{d \mathcal{E}}{dt} \Big)_{iz}^{soft} = - \frac{n_H e^4}{4\pi \varepsilon_0^2 mv} \ln \Lambda_{iz}^{soft} \label{eq:st_iz_form1}.
\end{eqnarray}
The stopping cross section $\sigma_{iz}^{sti, soft}(T)$ also measures the net effect of soft ionizing collisions on energy loss,
\begin{equation}
    \sigma_{iz}^{sti, soft}(T) = \int_0^{W_{bnd}} (B+W) \frac{d\sigma_{iz}(W, T)}{dW} dW
\end{equation}
where $W_{bnd} = B w_{bnd}$. After straight forward calculation, for $w_{bnd}=ht-1$,
\begin{equation}
\begin{split}
\sigma_{iz}^{sti, soft}&(T) = \frac{4\pi a_0^2 \alpha^4 N}{(\beta_t^2 + \beta_u^2 + \beta_b^2)2b'} \{ (f(ht-1) - f(0)) [\ln (\frac{\beta_t^2}{1-\beta_t^2}) -\beta_t^2 - \ln (2b')] \\
&+ (2-\frac{N_i}{N})[\frac{ht}{1+(1-h)t} -\frac{1}{t} + \ln \frac{1+(1-h)t}{t} \frac{1+2t'}{(1+t'/2)^2} + \ln ht \\
&+ \ln \frac{1+(1-h)t}{t} +\frac{b'^2}{(1+t'/2)^2}\frac{h^2t^2 - 1}{2}]  \},
\end{split}
\end{equation}
for $w_{bnd} = \frac{t-1}{2}$
\begin{equation}
\begin{split}
\sigma_{iz}^{sti, soft}&(T) = \frac{4\pi a_0^2 \alpha^4 N}{(\beta_t^2 + \beta_u^2 + \beta_b^2)2b'}\{ (f(\frac{t-1}{2}) - f(0)) [\ln (\frac{\beta_t^2}{1-\beta_t^2}) -\beta_t^2 - \ln (2b')] \\
&+(2-\frac{N_i}{N})[1-\frac{1}{t} + \ln \frac{t+1}{2t} \frac{1+2t'}{(1+t'/2)^2} + 2\ln \frac{t+1}{2} - \ln t +\frac{b'^2}{(1+t'/2)^2}\frac{t^2+2t-3}{8}]  \}
\end{split}
\end{equation}
and for $w_{bnd}=0$,
\begin{equation}
\begin{split}
\sigma_{iz}^{sti, soft}&(T) = 0
\end{split}
\end{equation}
where $f(w) = \int df/dw(w) dw = -b/(1+w) - c/(2(1+w)^2) - d/(3(1+w)^3) - e/(4(1+w)^4)$ with $b=-2.2473\cdot10^{-2}$, $c=1.1775$, $d=-4.6264\cdot10^{-1}$ and $e=8.9064\cdot10^{-2}$.

The stopping power estimated from $\sigma_{iz}^{sti, soft}(T)$ is
\begin{eqnarray}
    \Big( \frac{d \mathcal{E}}{dt} \Big)_{iz}^{soft} = - n_H v \sigma_{iz}^{sti,soft} \equiv - \nu_{iz}^{sti, soft} \label{eq:st_iz_form2}.
\end{eqnarray}
Equating Eqns. (\ref{eq:st_iz_form1}) and (\ref{eq:st_iz_form2}) yields
\begin{eqnarray}
    \ln \Lambda_{iz}^{soft} = \frac{4 \pi \varepsilon_0^2 mv}{n_H e^4} \nu_{iz}^{sti, soft}.
\end{eqnarray}

\subsubsection{Soft exciting collisions}
Let the stopping power of soft exciting collisions on electrons have the logarithmic factor. 
\begin{eqnarray}
    \Big( \frac{d \mathcal{E}}{dt} \Big)_{ex}^{soft} = - \frac{n_H e^4}{4\pi \varepsilon_0^2 mv} \ln \Lambda_{ex}^{soft} \label{eq:st_ex_form1}.
\end{eqnarray}
The stopping cross section of soft exciting collision is $\sigma_{ex}^{ste,soft} = \sum_{n=2,10} \sigma_{ex}^{1n} E_{1n} \mathcal{H}(T- \frac{E_{1n}}{h})$ due to fixed excitation energy. The stopping power estimated from $\sigma_{ex}^{ste, soft}(T)$ is
\begin{eqnarray}
    \Big( \frac{d \mathcal{E}}{dt} \Big)_{ex}^{soft} = - n_H v \sigma_{ex}^{ste,soft}  \equiv - \nu_{ex}^{ste, soft} \label{eq:st_ex_form2}.
\end{eqnarray}
Equating Eqns. (\ref{eq:st_ex_form1}) and (\ref{eq:st_ex_form2}) yields
\begin{eqnarray}
    \ln \Lambda_{ex}^{soft} = \frac{4 \pi \varepsilon_0^2 mv}{n_H e^4} \nu_{ex}^{ste, soft}
\end{eqnarray}

\subsubsection{Logarithmic factor}
The logarithm factor due to soft inelastic collisions is
\begin{equation}
    \ln \Lambda^{soft} = \frac{4 \pi \varepsilon_0^2 mv}{n_H e^4} (\nu_{iz}^{sti, soft} + \nu_{ex}^{ste, soft}) \label{eq:LnL_soft}.
\end{equation}
Remind that we deduce Eqn. \ref{eq:LnL_soft} by matching a function form of stopping power. That is to say, it does not originate from diverging nature of scattering cross sections in small energy transfer limit like free-free Coulomb collisions \citep{Rosenbluth1957PR}.

In the $h \to 1$ limit, all collisions are soft. The logarithmic factor due to total inelastic collisions can be obtained by taking this limit, i.e. $\ln \Lambda^{tot} = \lim_{h\to1} \ln \Lambda^{soft}$, since the test particle operator in the FPB is approximated to the FP. The hard collision part becomes $\ln \Lambda^{hard} = \ln \Lambda^{tot} - \ln \Lambda^{soft}$.

\begin{figure}
    \centering
    \includegraphics[width=0.8\linewidth]{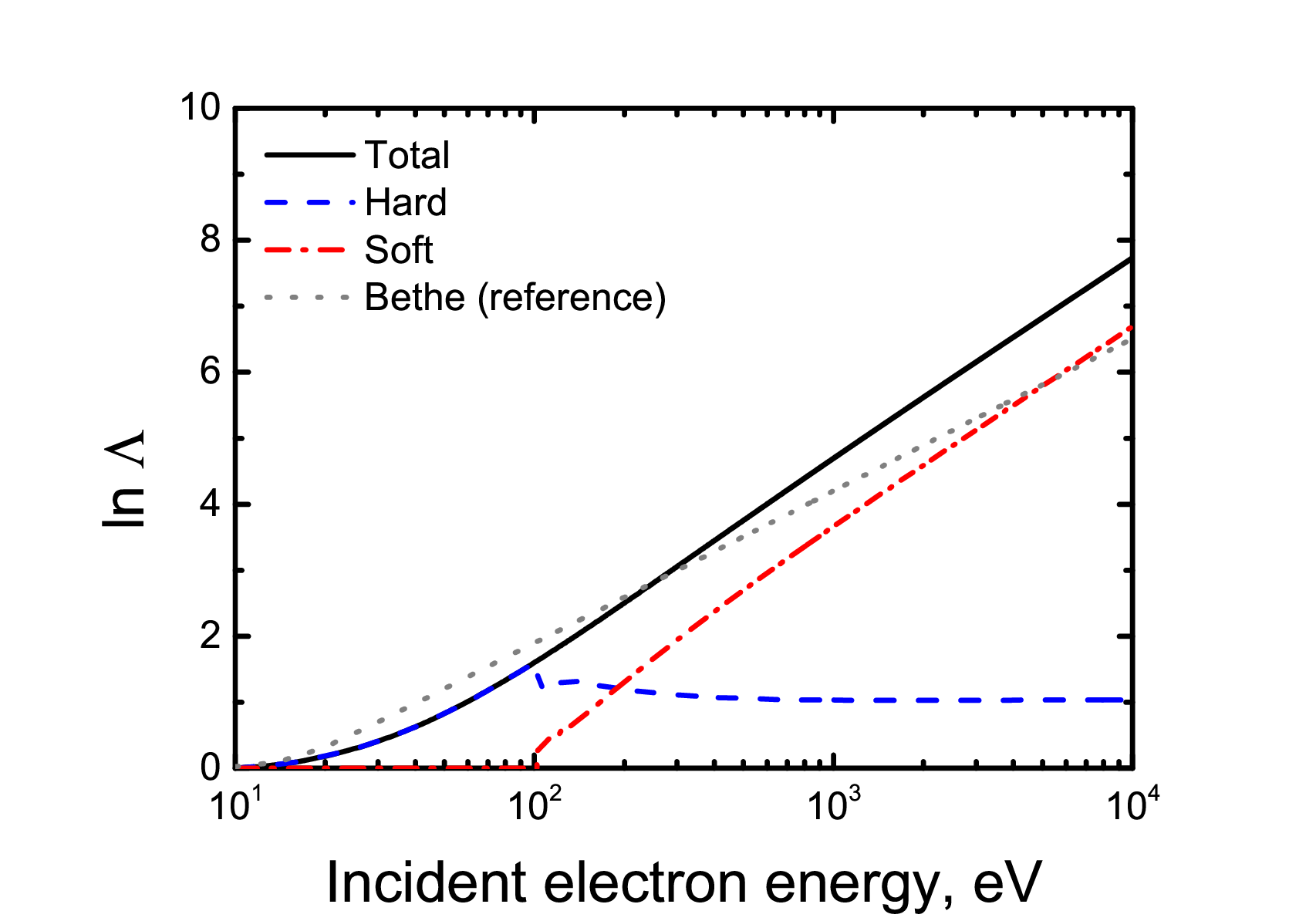}
    \caption{The logarithmic factor due to inelastic collisions, total (solid black), hard (blue dashed) and soft (red dashed-dotted) with $h=0.1$. Grey dotted line is the Bethe model extrapolated as reference.}
    \label{fig2}
\end{figure}

Figure \ref{fig2} shows smallness of the logarithmic factor for free-bound collisions. In the fully-ionized plasmas, the Coulomb logarithm measures the dominance of small angle scattering ($\ln \Lambda \gg 1$) and justifies neglect of higher order terms during derivation of the FP operator. In other word, such system meets $(\frac{\partial f}{\partial t})_c \approx - \vec{\nabla} \cdot (f \langle \Delta \vec{v} \rangle) + \frac{1}{2} \vec{\nabla}\vec{\nabla} : (f \langle \Delta \vec{v} \Delta \vec{v} \rangle)$ since high order terms like $\langle \Delta \vec{v} \Delta \vec{v} \Delta \vec{v} \rangle$ doesn't contain $\ln \Lambda$. For inelastic collisions, however, the logarithmic factor is so small that the assumption required for the expansion is broken. In addition, the inelastic collisions are predominantly hard for low energy particles. Such collisions need to be accounted by Eqn. (\ref{eq:hard}).

\subsubsection{Normalization}
\begin{figure}
    \centering
    \includegraphics[width=0.8\linewidth]{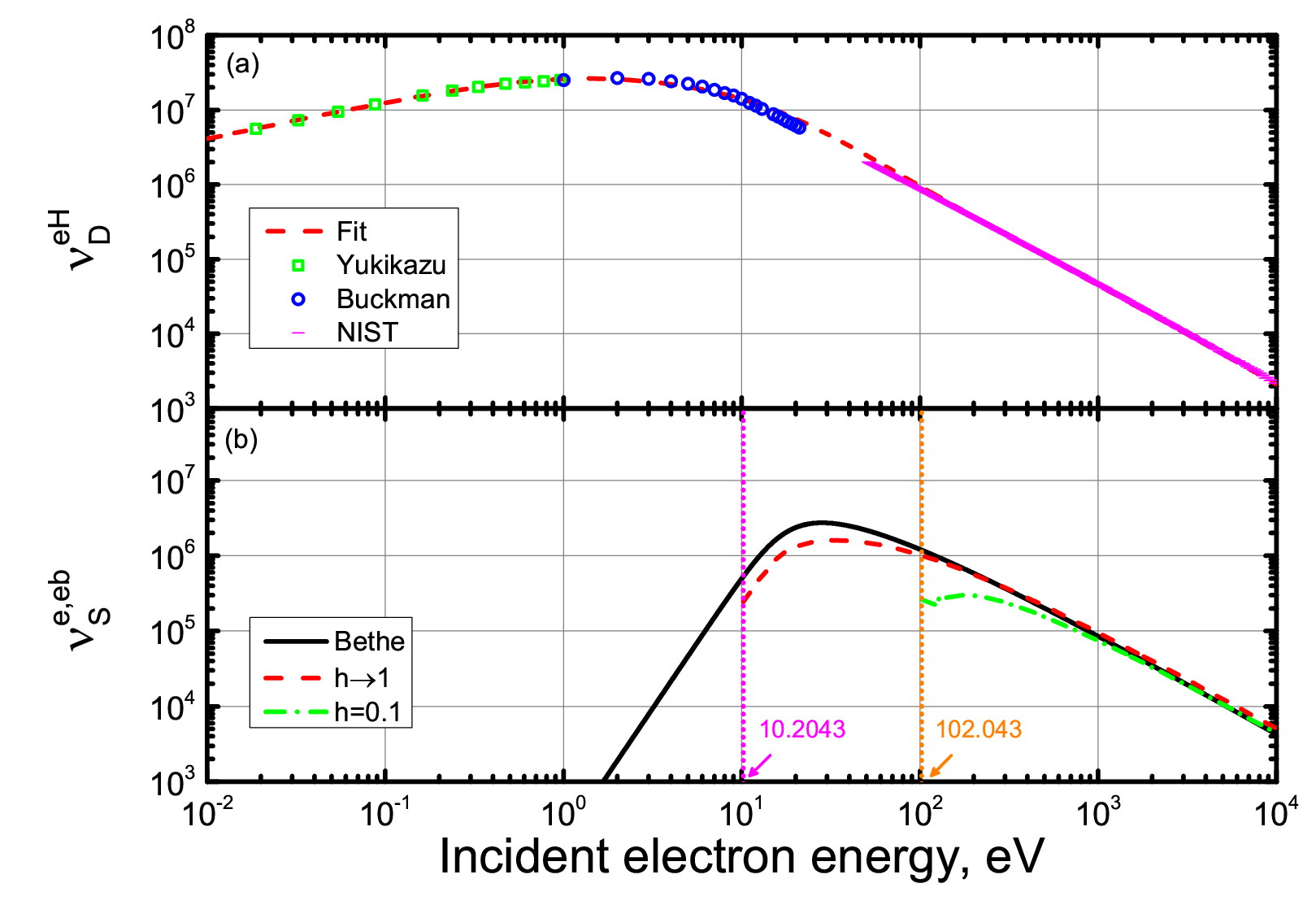}
    \caption{(a) The normalized deflection frequency and (b) slowing-down frequency as a function of incident electron energy. The legend of (a) is same to Fig. \ref{fig1}. (b) is from the extrapolated Bethe model \citep{Hesslow2017PRL} (black solid), the FPB model in $h \rightarrow 1$ limit (red dashed) and with $h=0.1$ (green dashed dotted). $n_e=10^{16} \ m^{-3}$ and $n_H=0.99 \cdot 10^{18} \ m^{-3}$. Reprinted figure with permission from \cite{Lee2024PRL}, copyright 2025 by the American Physical Society.}
    \label{fig3}
    \begin{subfigure}{0pt}
    \phantomcaption
    \label{fig3a}
    \end{subfigure}
    \begin{subfigure}{0pt}
    \phantomcaption
    \label{fig3b}
    \end{subfigure}
\end{figure}
Let $\tilde{\nu}_{iz}^{sti, soft} = \nu_{iz}^{sti, soft} \tau_c / mc^2$, $\tilde{\nu}_{ex}^{ste, soft} = \nu_{ex}^{ste, soft} \tau_c / mc^2$ and $\ln \tilde{\Lambda}^{soft} = n_H \ln \Lambda^{soft} /$ $ n_e \ln \Lambda$. Neglecting tilde, Then, the normalized logarithmic factor due to inelastic collisions $\ln \Lambda_{bound}$ is
\begin{equation}
    \ln \Lambda^{soft} = \beta (\nu_{iz}^{sti, soft} + \nu_{ex}^{ste, soft}).
\end{equation}

For $\mathcal{C}_{FP}^{soft} \{ f_e \}$, we took a form of Ref \citep{Helander2005Cam}. Using this form and Eqns. (\ref{eq:I_2}, \ref{eq:chi}, \ref{eq:LnL_soft}),  $2\pi p^2 \mathcal{C}_{FP}^{soft} \{ f_e \}$ becomes
\begin{equation}
    2\pi p^2 \mathcal{C}_{FP}^{soft} \{ f_e \} = \frac{\partial}{\partial p} p \nu_S^{e,eb} F_e + (\nu_D^{e,eb} + \nu_D^{eH}) \mathcal{L}\{F_e\}
\end{equation}
where $\mathcal{L}$ is the Laplace operator \citep{Helander2005Cam} and characteristic collisional frequencies are
\begin{eqnarray}
    \nu_S^{e,eb} = \frac{\gamma^2}{p^3} \ln \Lambda^{soft}, \\
    \nu_D^{e,eb} = \frac{\gamma}{p^3} \ln \Lambda^{soft}, \\
    \nu_D^{eH} = \frac{\gamma}{p^3} \chi \frac{n_H I_2^H(y)}{n_e \ln \Lambda}.
\end{eqnarray}

$\nu_D^{eH}$ and $\nu_S^{e,eb}$ are shown in Figs. \ref{fig3a} and \ref{fig3b}, respectively. All elastic collisions are accounted by $\nu_D^{eH}$ in $C_{FP}^{soft}$. For inealstic collisions of low energy electrons, the soft collisions represent only a fraction of all collisions and the associated frequency is therefore lower. In $h=0.1$ case, for instance, $\nu_S^{e,eb}=0$ in $\mathcal{C}^{eH}$ and $\mathcal{B}^{eH}$ treats all inelastic collisions below $102.043 \ eV$, this energy corresponds to $1/h$ of the minimal excitation energy for H. The ``continuous'' soft collision energy loss is significantly reduced even above $102.043 \ eV$. Nevertheless, the green curve goes to the red curve for high energy electrons due to the dominating soft collisions over the hard collisions.

\subsection{Invariance in $h$ \label{sec2-3}}
The collision operator $\mathcal{C}^{eH}$ is invariant in the total particle source rate and total energy loss rate with respect to $h$, i.e. $\frac{\partial}{\partial h} [ \int dp d\mu 2\pi p^2 \ \mathcal{C} \{ f_e \} ] = \frac{\partial}{\partial h} [ \int dp d\mu \ (\gamma-1) 2\pi p^2 \mathcal{C} \{ f_e \} ] = 0$. The invariance ensures the particle and energy conservation and, for a given distribution function, decouple an effect of energy loss allocation on electron acceleration from that of total energy loss.%This means any difference in electron acceleration originates from how we distribute the total energy loss to particles.

\subsubsection{Particle loss rate}
Let's take an integral $\int dp d\mu$ on $2\pi p^2 \mathcal{C}^{eH} \{ f_e \}= 2\pi p^2 \mathcal{C}^{soft}_{FP} \{ f_e \} + 2\pi p^2 \mathcal{B}^{hard}_{iz} \{ f_e \} + 2\pi p^2 \mathcal{B}^{hard}_{ex} \{ f_e \}$. The term $\int dp d\mu 2\pi p^2 \mathcal{C}^{soft}_{FP} \{ f_e \}$ automatically goes to $0$ due to its convective-diffusive form. An integral $\int d\mu$ on $2\pi p^2 \mathcal{B}^{hard}_{iz} \{ f_e \}$ yields
\begin{align}
    \int d\mu 2\pi p^2 \mathcal{B}^{hard}_{iz} \{ f_e \} &=  n_H mc^3 \tau_c \frac{p}{\gamma}\Big[ \int_{p_{bnd}} dp_1 \frac{\partial \sigma_{iz}}{\partial W}  \frac{p_1}{\gamma_1} F_e^0(p_1) \Big] - \nu_{iz}^{hard} F_e^0(p) \label{eq:int_mu_ptl_iz}
\end{align}
For an integral $\int dp$, change an integral variable from $p$ to $W$ and swap the order of integration, i.e. $\int dp \int_{p_{bnd}}dp_1 = \frac{1}{mc^2}\int dp_1 \int_0^{T-B-W_{bnd}} dW$. Then an integral part $\int_0^{T-B-W_{bnd}} dW \frac{\partial \sigma_{iz}^{hard}}{  \partial W} = \sigma_{iz}+\sigma_{iz}^{hard}$ is isolated. Outcome of taking an integral $\int dp$ on Eqn. (\ref{eq:int_mu_ptl_iz}) is independent of $h$, i.e.
\begin{align}
    \int dp d\mu 2\pi p^2 \mathcal{B}^{hard}_{iz} \{ f_e \} &= \int dp_1 \ (\nu_{iz} + \nu_{iz}^{hard}) F_e^0(p_1) - \int dp \ \nu_{iz}^{hard} F_e^0(p) \notag \\
    &= \int dp \ \nu_{iz} F_e^0(p).
\end{align}

Using $\beta(p)dp = \beta(p^+_{1n})dp^+_{1n}$, the excitation part is straightforward.
\begin{align}
    \int dp d\mu 2\pi p^2 \mathcal{B}^{hard}_{ex, 1s \to np} &\{ f_e \} = \int dp^+_{1n} \Big[ \nu_{ex,1s \to np}^{hard}(p^+_{1n}) F_e^0(p^+_{1n}) \Big] \notag \\
    &- \int dp \Big[ \nu_{ex,1s \to np}^{hard}(p) F_e^0(p) \Big] = 0
\end{align}
Finally, the outcome,
\begin{equation}
    \int dp d\mu 2\pi p^2 \mathcal{C}^{eH} \{ f_e \}=\int dp \ \nu_{iz} F_e^0(p) \label{eq:int_fin_ptl}
\end{equation}
suggests the total particle loss is invariant in $h$.

\subsubsection{Energy loss rate}
Take an integral $\int dp d\mu d\gamma$ on $2\pi p^2 \mathcal{C}_{FP}^{soft}$ and then the pitch angle scattering part automatically goes to $0$. The remaining part is
\begin{align}
    \int dp \gamma \partial_p (p \nu_S^{e,eb} F_e^0(p)) &= \int dp  \Big[ -\nu_{iz}^{sti,soft}(T) \Big] F_e^0(p) + \int dp \Big[ - \nu_{ex}^{ste,soft}(T) \Big] F_e^0(p) \label{eq:int_fin_ene_FP}.
\end{align}

Take an integral $\int d\mu$ on ionization part in the Boltzmann operator to eliminate the scattering angle term.
\begin{align}
    \int d\mu 2\pi p^2 \mathcal{B}_{iz}^{hard} &= n_H mc^3 \tau_c \frac{p}{\gamma} \int_{p_{bnd}} dp_1 \Big[\frac{\partial \sigma_{iz}(W, T)}{\partial W} \frac{p_1}{\gamma_1} F_e^0 (p_1)\Big] - \nu_{iz}^{hard} F_e^0 (p) \label{eq:int_mu_ene_iz}
\end{align}
For an integral $\int \gamma dp$, change an integral variable from $p$ to $W$ and swap the order of integration, i.e. $\int dp \int_{p_{bnd}}dp_1 = \frac{1}{mc^2}\int dp_1 \int_0^{T-B-W_{bnd}} dW$. This isolates integral of $\frac{\partial \sigma_{iz}(W, T)}{\partial W}$. From $0$ to $(T-B)/2$, the result of the integral is $\int_{0}^{(T-B)/2} dW \gamma \frac{\partial \sigma_{iz}(W, T)}{\partial W}= \frac{1}{mc^2}\sigma_{iz}^{sti} + (1-b') \sigma_{iz}$. Then, take a variable transformation from $W$ to $T-B-\tilde{W}$ on the remaining part.
\begin{align}
    \int_{0}^{T-B-W_{bnd}} &dW \gamma \frac{\partial \sigma_{iz}(W, T)}{\partial W} = \frac{1}{mc^2} \sigma_{iz}^{sti} + (1-b') \sigma_{iz} \notag \\
    &+ \int_{W_{bnd}}^{(T-B)/2} d\tilde{W} \gamma(T-B-\tilde{W}) \frac{\partial \sigma_{iz}(\tilde{W}, T)}{\partial \tilde{W}}
\end{align}
The outcome of $\int dp \gamma$ on Eqn. (\ref{eq:int_mu_ene_iz}), resolving $\nu^{hard}_{iz}= n_H c \tau_c \frac{p}{\gamma} \int_{W_{bnd}}^{(T-B)/2} d\tilde{W} \frac{\partial \sigma_{iz}(\tilde{W}, T)}{\partial \tilde{W}}$, becomes
\begin{equation}
\begin{split}
    \int d\mu dp \gamma 2\pi p^2 \mathcal{B}_{iz}^{hard} &= \int dp_1 \Big[ \nu_{iz}^{sti} + (1-b') \nu_{iz} \Big] F_e^0(p_1)    
    + n_H c \tau_c \int dp \frac{p}{\gamma} F_e^0 (p) \\
    &\times [\int_{W_{bnd}}^{(T-B)/2} d\tilde{W} (\gamma(T-B-\tilde{W}) - \gamma(T)) \frac{\partial \sigma_{iz}(\tilde{W}, T)}{\partial \tilde{W}}] \\ 
    &= \int dp \Big[ \nu_{iz}^{sti, soft} + (1-b') \nu_{iz} \Big] F_e^0(p) \label{eq:int_fin_ene_iz}.
\end{split}
\end{equation}

\begin{figure*}
\centering
\begin{subfigure}[b]{0.48\textwidth}
    \centering
    \includegraphics[width=\textwidth]{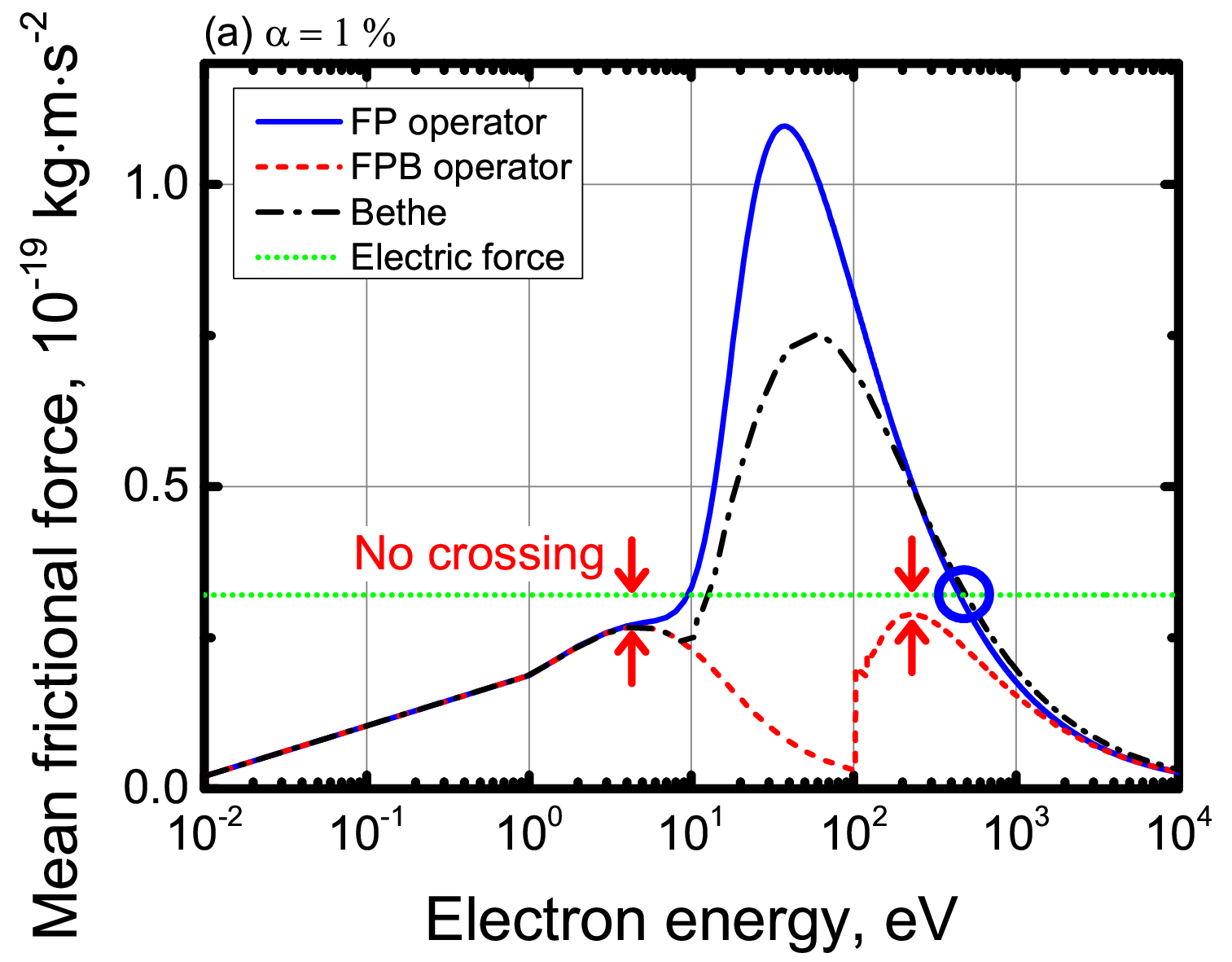}
    \phantomcaption
    \label{fig4a}
\end{subfigure}
\hfill
\begin{subfigure}[b]{0.48\textwidth}
    \centering
    \includegraphics[width=\textwidth]{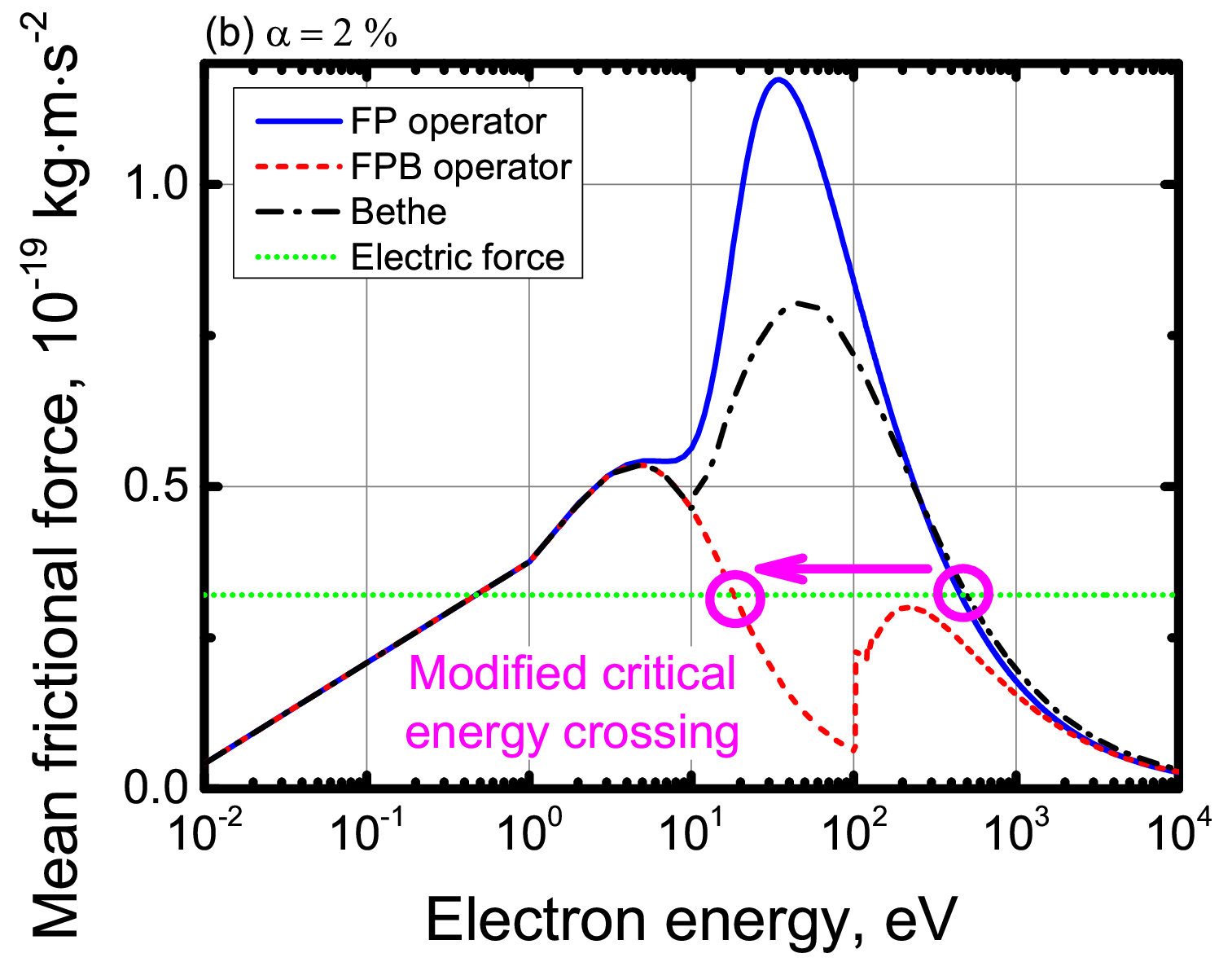}
    \phantomcaption
    \label{fig4b}
\end{subfigure}
\\
\begin{subfigure}[b]{0.48\textwidth}
    \centering
    \includegraphics[width=\textwidth]{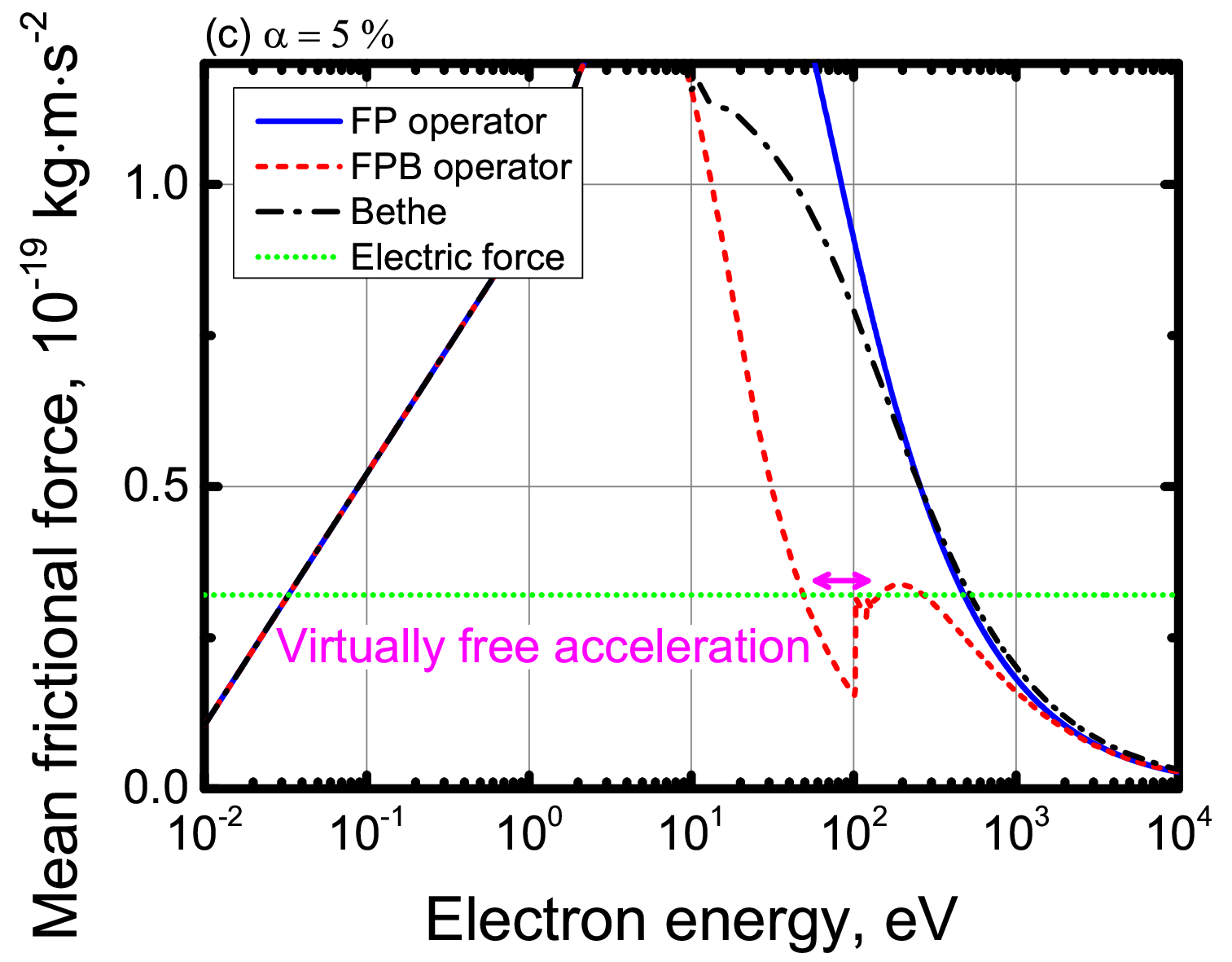}
    \phantomcaption
    \label{fig4c}
\end{subfigure}
\hfill
\begin{subfigure}[b]{0.48\textwidth}
    \centering
    \includegraphics[width=\textwidth]{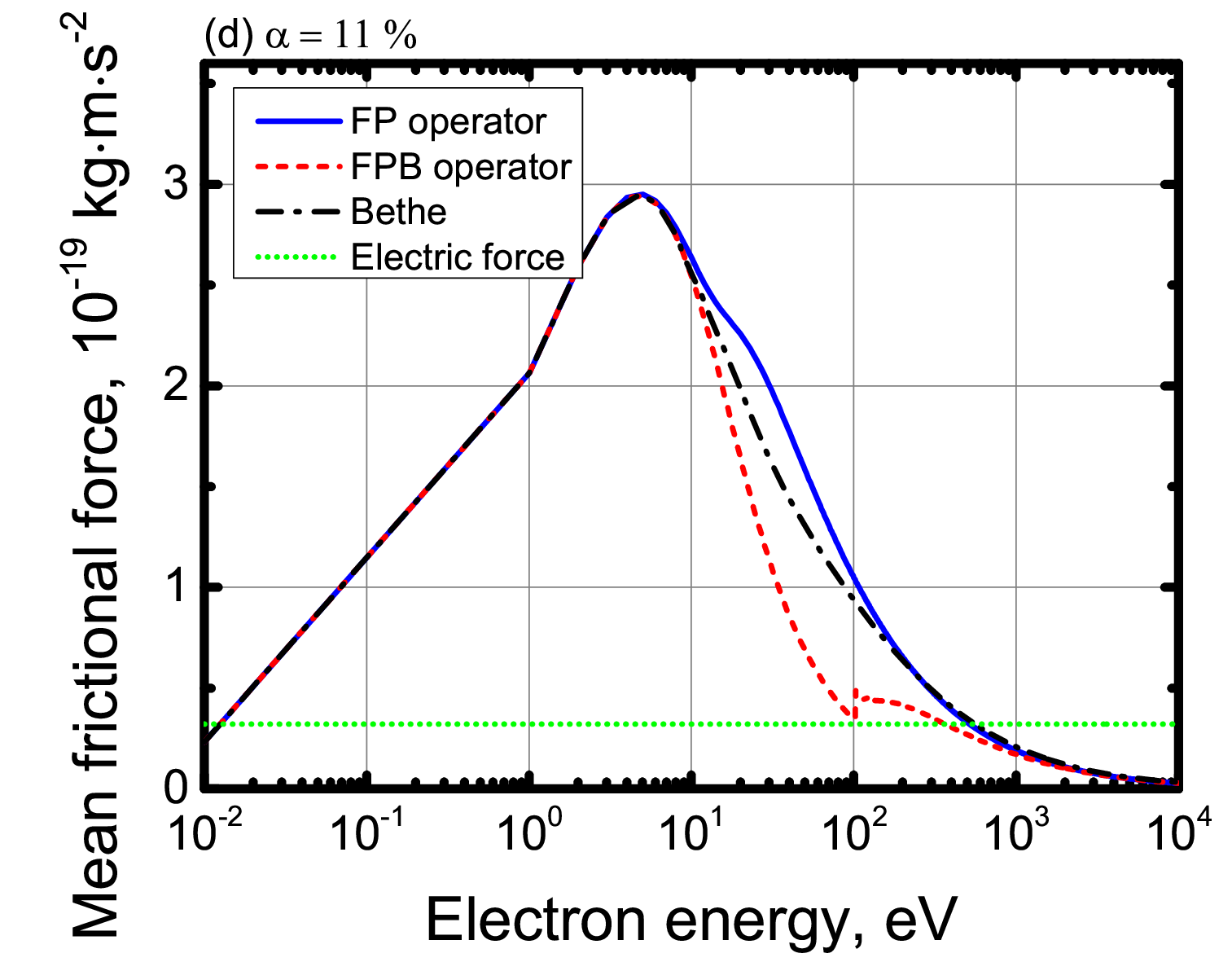}
    \phantomcaption
    \label{fig4d}
\end{subfigure}
\caption{Mean frictional force considered by the FP operator as a function of incident electron energy. Free-bound collisions are from the extrapolated Bethe model \cite{Hesslow2017PRL} (black dashed dotted curve), the FPB model in $h \rightarrow 1$ limit, same to the FP model, (blue solid curve) and with $h=0.1$ (red dashed curve). Green dotted curve is the electric force. $n_e + n_H =10^{18} \ m^{-3}$, $E=0.2 \ Vm^{-1}$, $T_e = 5 \ eV$ and $\ln \Lambda = 15$. $\alpha = 1, \ 2, \ 5$ and $11 \ \%$ for (a), (b), (c) and (d), respectively.}
\label{fig4}
\end{figure*}

Take a $\int d\mu$ on excitation part ($1s \to np$ transition) in the Boltzmann operator. 
\begin{align}
    \int d\mu 2\pi p^2 \mathcal{B}_{ex, 1s \to np}^{hard} &= \nu_{ex, 1s \to np}^{hard}(p^+_{1n}) \frac{\beta(p)}{\beta(p^+_{1n})} F_e^0 (p^+_{1n}) - \nu_{ex, 1s \to np}^{hard}(p) F_e^0 (p)
\end{align}
For $\int dp \gamma$, use $\beta(p^+_{1n})dp^+_{1n} = \beta(p) dp$,
\begin{align}
    \int d\mu dp \gamma 2\pi p^2 \mathcal{B}_{ex, 1s \to np}^{hard} &=  \int dp^+_{1n}\gamma \Big[\nu_{ex, 1s \to np}^{hard}(p^+_{1n}) F_e^0 (p^+_{1n})\Big] - \int dp\gamma \Big[\nu_{ex, 1s \to np}^{hard}(p) F_e^0 (p) \Big] \notag \\
    &= \int dp \Big[(\gamma(T-E_{1n}) - \gamma(T)) \Big]\nu_{ex, 1s \to np}^{hard} F_e^0 (p) \notag \\
    &= \int dp \Big[ -\nu_{ex, 1s \to np}^{ste, hard} \Big] F_e^0 (p)  \label{eq:int_fin_ene_ex}
\end{align}
Finally, Eqns. (\ref{eq:int_fin_ptl}, \ref{eq:int_fin_ene_FP}, \ref{eq:int_fin_ene_iz} and \ref{eq:int_fin_ene_ex})) yield the outcome,
\begin{equation}
    \int dp d\mu (\gamma-1) 2\pi p^2 \mathcal{C}^{eH} \{ f_e \}=\int dp \ \Big[ - b'\nu_{iz} - \nu_{ex}^{ste} \Big] F_e^0(p)
\end{equation}
and suggests the total energy loss is invariant in $h$.

\subsection{Collisions with free electrons and ions \label{sec2-4}}
We describe the kinetic model for collisions with free electrons and ions for completeness of this paper. The majority of free electrons are assumed to be Maxwellian as \textit{a posteriori}. The justification is essential for applying it to weakly-ionized plasmas and hence will be verified through numerical simulation in Sec. \ref{sec:2D4}.

The FP operator is used to account for the Coulomb collisions with free electrons and ions:
\begin{equation}
    2\pi p^2 \mathcal{C}_{FP} \{ f_e \} = 2\pi p^2 \mathcal{C}_{FP}^{e,i} \{ f_e \} + 2\pi p^2 \mathcal{C}_{FP}^{e,ef} \{ f_e \}.
\end{equation}
The collisions with ions are described by the Lorentz operator \cite{Helander2005Cam}:
\begin{equation}
    2\pi p^2 \mathcal{C}_{FP}^{e,i} \{ f_e \} = \nu_D^{ei} \mathcal{L}\{F_e\}.
\end{equation}
The deflection frequency is
\begin{equation}
    \nu_D^{ei} = \frac{\gamma}{p^3} \frac{ \ln \Lambda_{free} }{\ln \Lambda},
\end{equation}
where $\ln \Lambda_{free}$ is in Ref. \citep{Breizman2019NF}.

For collisions with free electrons, we linearize $\mathcal{C}_{FP}^{e,ef}$ due to negligible order of nonlinear terms:
\begin{equation}
    2\pi p^2 \mathcal{C}_{FP}^{e,ef} \{ f_e \} \approx 2\pi p^2 \mathcal{C}_{FP}^{e,ef} \{ f_e, f_M \} + 2\pi p^2 \mathcal{C}_{FP}^{e,ef} \{ f_M, f_e \} \label{eq:FP_full}
\end{equation}
where $f_M$ is the Maxwell distribution.

\subsubsection{Test particle part}
The test particle operator has a form of Ref. \citep{Papp2011NF}:
\begin{equation}
\begin{split}
    2\pi p^2 \mathcal{C}_{FP} \{ f_e, f_M \} &= \frac{\partial}{\partial p} \Big[ p \nu_S^{e,ef} - \frac{\nu_\parallel^{e,ef}}{p} + \frac{1}{2} p^2 \nu_\parallel^{e,ef} \frac{\partial}{\partial p} \Big] F_e + (\nu_D^{e,ef} + \nu_D^{ei}) \mathcal{L}\{F_e\} \label{eq:FP_test}.
\end{split}
\end{equation}
This operator is suitable for arbitrary electron energy. 
The collision frequencies are
\begin{eqnarray}
    \nu_S^{e,ef} &=& \frac{G(X_G)}{\bar{T}_e p} \frac{\ln \Lambda_{free}}{\ln \Lambda}, \\
    \nu_\parallel^{e,ef} &=& 2\frac{\gamma}{p^3} G(X_G) \frac{\ln \Lambda_{free}}{\ln \Lambda}, \\
    \nu_D^{e,ef} &=& \frac{\gamma}{p^3}(\Phi(X_G) - G(X_G) + \bar{T}_e\beta^2)\frac{\ln \Lambda_{free}}{\ln \Lambda} \label{eq:nuD_eef}, \\
    \nu_D^{ei} &=& \frac{\gamma}{p^3} \frac{ \ln \Lambda_{free} }{\ln \Lambda},
\end{eqnarray}
where $\bar{T}_e = T_e / mc^2$, $X_G = v/v_{th}$ is the velocity normalized by the thermal velocity $v_{th} = \sqrt{2 T_e / m}$, $G(X_G)$ is the Chandraseckhar function, $\Phi(X_G)$ is the error function and $\ln \Lambda_{free}$ is in Ref. \cite{Breizman2019NF}.

\subsubsection{Field particle part}
The field particle operator is necessary to explicitly justify the background Maxwell assumption provided that the exponential factor is not negligible at the core.

The field particle operator has a form of Ref. \citep{Li2011PRL}:
\begin{equation}
\begin{split}
    2\pi p^2 \mathcal{C} & \{ f_M, f_e \} = \frac{1}{2\pi} \int dp' d\mu' \Big[\frac{1}{\beta_T^4} \Big((\beta^2 + \beta'^2)\frac{4 K(\kappa)}{\lambda} - 2 \lambda E(\kappa) - (\beta^2 - \beta'^2)^2 \frac{2E(\kappa)}{\lambda^3 (1-\kappa^2)}\Big) \\
    &-\frac{2}{\beta_T^2} \frac{4 K(\kappa)}{\lambda} \Big] \bar{F}_M(\vec{p}) F_e (\vec{p}') + \frac{2}{p^2} \bar{F}_M (\vec{p}) F_e (\vec{p}) \label{eq:FP_field}
\end{split}
\end{equation}
where $\bar{F}_M = 2 \pi p^2 f_M / n_e$, $\beta_T = v_T/c$, $\lambda^2 = (\beta_\perp + \beta'_\perp)^2 + (\beta_\parallel - \beta'_\parallel)^2$ with $\beta_\perp = \beta \sqrt{1-\mu^2}$ and $\beta_\parallel = \beta \mu$ and $\kappa^2 = 4 \beta_\perp \beta'_\perp / \lambda^2$. $K$ and $E$ are the complete elliptic integral of the first and second kinds, respectively.

\subsubsection{Conservation laws}
The conservation laws are approximately satisfied with the free-free collision operator Eqn. (\ref{eq:FP_full}). The relativistic operator that was matched to obtain Eqn. (\ref{eq:FP_test}) \citep{Papp2011NF} originates from the Beliaev-Budker kernel \citep{Beliaev1956, Landau1981} by taking the Taylor expansion upto the 2nd order \citep{Sandquist2006POP}. However, the only correction given by this matching from the semi-relativistic operator \citep{Helander2005Cam} is the $\bar{T}_e \beta^2$ term in Eqn. (\ref{eq:nuD_eef}) that is much smaller than $\Phi(X_G) \to 1$ for energetic electrons due to $\bar{T}_e \ll 1$. Therefore, both operators can be represented by the Landau kernel \citep{Landau1981} and conservation laws are approximately satisfied \citep{Landau1981} in cold weakly-ionized plasmas. Note that the Landau kernel is indeed semi-relativistic and appropriate if one of colliding particles is non-relativistic \citep{Braams1987PRL}.

\subsubsection{Verification of the Maxwellian assumption \label{sec:2D4}}
\begin{figure}
    \centering
    \includegraphics[width=0.8\linewidth]{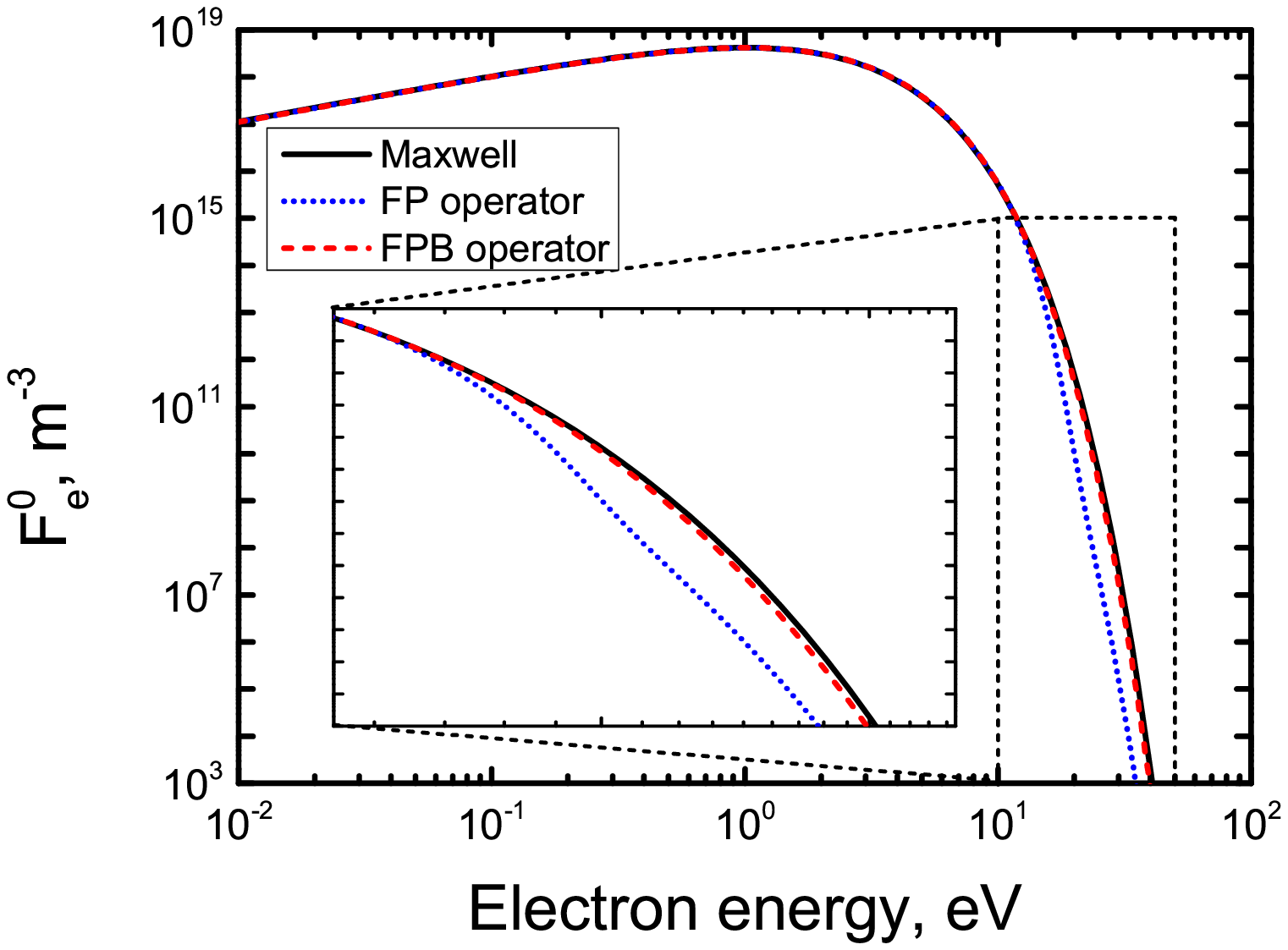}
    \caption{The Maxwell distribution function (black solid curve) and $F_e^0$ that evolves for $20 \ \mu s$ from the black curve with the FPB (red dashed curve, h=0.1) and FP (blue dotted curve, h=0.99) operators. $(n_H + n_e) = 10^{18} \ m^{-3}$, $T_e  =  1 \ eV$ and $\alpha = 1 \ \%$.}
    \label{fig5}
\end{figure}
The "core" Maxwell assumption can be guaranteed if $n_e \ln \Lambda_{free} \gg n_H \ln \Lambda_{bound}$, where $\ln \Lambda_{free}$ is the Coulomb logarithm for free-free collisions and $\ln \Lambda_{bound}$ is the logarithmic factor for free-bound collisions. This condition is evidently satisfied below $100 \ eV$ because $\ln \Lambda_{bound}$ felt by a majority of electrons is $\ln \Lambda^{soft}$ and thus is vanishingly small (see red curve in Fig. \ref{fig2}). Therefore, the lower energy part of the population is predominantly Maxwellian. The argument is only true for the background electron distribution, not for the entire electron distribution.

The adequacy of the background Maxwellian assumption is numerically verified by confirming whether the Maxwell distribution function is sustained or not under the effect of free-bound collisions accounted by $\mathcal{C}$. The numerical simulation is carried out by initiating $F_M$ and then evolving it for $20 \ \mu s$. Figure \ref{fig5} shows resulting $F_e^0$. The black curve aligns well with the red curve. This good agreement represents inelastic collisions don't distort $F_e^0$ as severely as violate the Maxwell assumption. In this specific example, a total friction of free-bound collision dominates free-free collision due to $n_e \ln \Lambda_{free} \ll n_H \ln \Lambda^{tot}$. Nonetheless, the Maxwell assumption is valid because a majority of electrons completely don't experience inelastic collisions, i.e. $n_e \ln \Lambda_{free} \gg n_H \ln \Lambda^{soft}$. Indeed, free-bound collisions are strong, as seen by the blue curve, which clearly deviates from Maxwellian due to the misuse of the FP operator. For the verification, the ionizing avalanche is regulated by adopting the test particle part in $\mathcal{C}^{eH}$ while the field particle part is included in $\mathcal{C}_{FP}$ for the low energy population. The detailed description for the implementation is in Sec. \ref{sec3}.

\subsection{Non-diffusive electron acceleration mechanism \label{sec2-5}}
The free-free and free-bound interaction exhibit significantly different interaction distances. This difference distinguishes the electron acceleration mechanisms. For free-free collisions, \textit{all} electrons feel the interactions due to the long-range nature in single particle description. In kinetic description, electrons with same energy have the \textit{same} slowing down frequency. The FP operator takes into account disparities in the momentum exchanges using the velocity space diffusion frequency. Through the diffusion mechanism within the narrow singular layer across the critical momentum in phase space, the upward Dreicer flow can form under the effect of electric field, below which the mean frictional force exceeds the electric force. In a contrast, a range of interaction is short for free-bound inelastic collisions due to a screened charge of neutrals in single particle description. Therefore, \textit{some} electrons can accelerate freely without undergoing inelastic collisions whenever they encounter a neutral particle. 
In kinetic description, % Electrons can share the slowing down frequency only if the associated energy transfer is small compared to their energy, i.e. for soft collisions. In low energy region, however, %\textit{some} electrons completely unparticipate in inelastic collisions and accelerate freely. 
this mechanism is evidently distinct with the diffusive process and occurs over the electrons’ acceleration trajectories in low energy region of phase space, where most inelastic collisions are hard so should be described by the Boltzmann operator; electrons can share the slowing down frequency only if the associated energy transfer is small compared to their energy, i.e. for soft collisions.

Such accounting of binary nature of inelastic collisions changes dramatically the critical energy crossing for free acceleration. This is clarified in Fig. \ref{fig4} that shows mean frictional forces considered by the FP operator. Note that the mean frictional force is shared by electrons with the same energy through the slowing down frequency in each model. Other collisions considered by the binary Boltzmann operator are excluded in this computation. If the FP operator describes all collisions (blue or black curve), electron runaway begins at the critical energy crossing (e.g. see blue circle in Fig. \ref{fig4a}). However, the FPB treatment can eliminate the crossing as shown in Fig. \ref{fig4a} if the ionization fraction is sufficiently low. In this case, the Dreicer condition \citep{Dreicer1959PR} is completely satisfied for some electrons: all of electrons that don't experience hard inelastic collisions are runaway. Even in less extreme cases, the proposed model significantly modifies the critical energy by an order of magnitude (in Fig. \ref{fig4b}) or allows the virtual free acceleration (in Fig. \ref{fig4c}). The increased ionization fraction makes the change in the critical energy less significant. Yet, the frictional force can be notably reduced even in this case (in Fig. \ref{fig4d}). In summary, the proposed model captures the non-diffusive Dreicer mechanism and allows for variations in the critical energy beyond the mean-friction approximation. Another noteworthy aspect is the associated diffusion with soft inelastic collisions, which will be addressed in a future study.

\section{Numerical implementation \label{sec3}}
We implement the FPB operator in the self-consistent kinetic simulation \citep{Aleynikov2017NF} using FiPy Finite Volume Partial Differential Equation (PDE) Solver \citep{Guyer2009CSE}. In FiPy, the discretization of $\mathcal{B}^{hard} \{ F_e \}$ is given by
\begin{equation}
    \int_{p_{l,i}}^{p_{r,i}} dp \int_{\mu_{l,j}}^{\mu_{r,j}} d\mu \mathcal{B}^{hard} \{ F_e \} \approx  \mathcal{B}^{hard} \{ F_e \} (p_i, \mu_j) dp_i d\mu_j,
\end{equation}
where $(p_i, \mu_j)$ is (i,j)-th cell center, $l$ and $r$ denote the left and right side cell face, respectively, $dp_i = p_{r,i}-p_{l,i}$ and $d\mu_j = \mu_{r,j}-\mu_{l,j}$.

\subsection{Discretization of $\mathcal{B}^{hard}_{iz} \{ F_e \}$}
Consider $\int_{\mu_{l,j}}^{\mu_{r,j}} d\mu $ on $\mathcal{B}^{hard}_{iz} \{ F_e \}$ and discretize $\mu$ integral using the relation,
\begin{align}
    \int_{\mu_{l,j}}^{\mu_{r,j}} d\mu \frac{\mathcal{H}((1-\mu^2)(1-\mu_1^2)-(\mu^*-\mu\mu_1)^2)}{\sqrt{(1-\mu^2)(1-\mu_1^2)-(\mu^*-\mu\mu_1)^2}} &= \sin^{-1} \Big[ \min_2 \Big(-1, \frac{\mu_{r,j} - \mu_1 \mu^*}{\sqrt{(1-\mu_1^2)(1-\mu^{*2})}}, 1 \Big) \Big] \notag \\
    &- \sin^{-1} \Big[ \min_2 \Big(-1, \frac{\mu_{l,j} - \mu_1 \mu^*}{\sqrt{(1-\mu_1^2)(1-\mu^{*2})}}, 1 \Big) \Big].
\end{align}
$\sum_j$ erases $\mu_j$ dependence owing to $\mu_{l,0}=-1$ and $\mu_{r,-1}=1$, where $-1$ index indicates the last one. Subsequently, discretize $p$ integral by taking $\int_{p_{l,i}}^{p_{r,i}} dp$ on $\sum_j \int_{\mu_{l,j}}^{\mu_{r,j}} d\mu \mathcal{B}^{hard}_{iz} \{ F_e \}$ and using the relation,
\begin{align}
    \int_{p_{l,i}}^{p_{r,i}} dp &\Big[ mc^2 \frac{p}{\gamma} \frac{\partial \sigma_{iz}(W,T)}{\partial W} \Big] = d\sigma_{iz}(w_{r,i}, T) - d\sigma_{iz}(w_{l,i}, T)
\end{align}
where $w_{r,i}= (\sqrt{p_{r,i}^2+1}-1)/b'$, $w_{l,i}= (\sqrt{p_{l,i}^2+1}-1)/b'$ and $d\sigma_{iz}(w,T)$ is
\begin{align}
    d\sigma_{iz}(w,T) &\equiv \int_{0}^{w}dw \frac{\partial \sigma_{iz}(W, T)}{\partial W} = \frac{4\pi a_0^2 \alpha^4 N}{(\beta_t^2 + \beta_u^2 + \beta_b^2)2b'} \{ \bar{D}(0, w) [\ln (\frac{\beta_t^2}{1-\beta_t^2}) -\beta_t^2 - \ln (2b')] \notag \\
    &+ (2-\frac{N_i}{N})[1 - \frac{1}{t} + \frac{1}{t-w} -\frac{1}{w+1}  -\ln((w+1)\frac{t}{t-w})\frac{1}{t+1} \frac{1+2t'}{(1+t'/2)^2} \notag \\
    &+\frac{b'^2}{(1+t'/2)^2}w] \}.
\end{align}
Here, $\bar{D}(0, w) = \frac{1}{N} \int_{0}^{w}\frac{1}{1+w}\frac{df}{dw}dw= \frac{b}{2}[1-\frac{1}{(1+w)^2}] + \frac{c}{3}[1-\frac{1}{(1+w)^3}] + \frac{d}{4}[1-\frac{1}{(1+w)^4}] + \frac{e}{5}[1-\frac{1}{(1+w)^5}]$. Swapping the order of integration with $p_1$, $\sum_i$ erases $p_i$ dependence:
\begin{align}
    \sum_i \int_{p_{pnd}} &dp_1 d\mu_1 \Big[ d\sigma_{iz}(w_{r,i}, T) - d\sigma_{iz}(w_{l,i}, T) \Big] \notag \\
    &= \int dp_1 d\mu_1 \sum_{i<i^*} \Big[ d\sigma_{iz}(w_{r,i}, T) - d\sigma_{iz}(w_{l,i}, T) \Big] \notag \\
    &= \int dp_1 d\mu_1 \Big[ \sigma_{iz}(T) + \sigma_{iz}^{hard}(T) \Big]
\end{align}
where we replace $d\sigma_{iz}(w_{l,0}, T)$ as $d\sigma_{iz}(0, T)$, $d\sigma_{iz}(w_{r,i^*}, T)$ as $d\sigma_{iz}(t-w_{bnd}-1, T)$ and $i^*$ satisfies $w_{l,i^*}<t-w_{bnd}-1$ and $w_{l,i^*+1}\ge t-w_{bnd}-1$. Then, the ionization part reduces to
\begin{align}
    &\sum_{i,j} \int_{p_{l,i}}^{p_{r,i}} dp \int_{\mu_{l,j}}^{\mu_{r,j}} d\mu \mathcal{B}^{hard} \{ F_e \} \notag \\
    &= \sum_{k,l'} \int_{p_{l,k}}^{p_{r,k}} \int_{\mu_{l,l'}}^{\mu_{r,l'}} \Big[(\nu_{iz}(T_k)+\nu_{iz}^{hard}(T_k)) F_e (p_k, \mu_{l'})\Big] dp_k d\mu_{l'} \notag \\
    &- \sum_{i,j} \int_{p_{l,i}}^{p_{r,i}} \int_{\mu_{l,j}}^{\mu_{r,j}} \Big[\nu_{iz}^{hard}(T_i)) F_e (p_i, \mu_j)\Big] dp_i d\mu_j.
\end{align}

We discretize $\int_{p_{l,k}}^{p_{r,k}} dp_k [\nu_{iz}(T_k)+\nu_{iz}^{hard}(T_k)) F_e (p_k, \mu_{l'})]$ grounded in the fact that $F_e(p_k,\mu_{l'})$ shows the exponential dependence on $p_k$ and both of $\nu_{iz}(T_k)$ and $\nu_{iz}^{hard}(T_k)$ have the algebraic dependence on $p_k$, i.e. $|\partial_p \ln (F_e)| \gg |\partial_p \ln(\nu_{iz})|$ and $|\partial_p \ln (F_e)| \gg |\partial_p \ln(\nu_{iz}^{hard})|$:
\begin{gather}
    \int_{p_{l,k}}^{p_{r,k}} dp_k \nu_{iz}(T_k) F_e (p_k, \mu_{l'}) \approx \nu_{iz}(T_k) F_e (p_k, \mu_{l'}) dp_k \label{eq:disc_pk_iztot} \\
    \int_{p_{l,k}}^{p_{r,k}} dp_k \nu_{iz}^{hard}(T_k) F_e (p_k, \mu_{l'}) \approx \nu_{iz}^{hard}(T_k) F_e (p_k, \mu_{l'}) dp_k \label{eq:disc_pk_izhard}
\end{gather}
An exception where the condition $|\partial_p \ln (F_e)| \gg |\partial_p \ln(\nu_{iz})|$ can be severely broken and influence the discretization is nearby $T \approx B$ due to onset of ionizing collisions. We replace $dp_{k^*}=p_{r,k^*}-p_{l,k^*}$ in Eqns. (\ref{eq:disc_pk_iztot} and \ref{eq:disc_pk_izhard}) as $dp_{k^*}=p_{r,k^*}-\sqrt{(b'+1)^2-1}$ where $k^*$ satisfies $T_{r,k^*} > B$ and $T_{l,k^*} < B$. The remaining $\mu_{l'}$ part is straightforward: $\int_{\mu_{l,l'}}^{\mu_{r,l'}} d\mu F_e(p,\mu) = F_e(p,\mu_{l'}) d\mu_{l'}$. Finally, the fully-discretized form of $\mathcal{B}^{hard}_{iz} \{ F_e \}$ is
\begin{align}
    [\mathcal{B}^{hard}_{iz} &\{ F_e \} (p_i, \mu_j)]_{k,l'} = \frac{1}{dp_i d\mu_j} \Big[ \Big\{ \frac{1}{\pi} \Big( \sin^{-1} [\min_2 (-1, \frac{\mu_{r,j}-\mu_1 \mu^*}{\sqrt{(1-\mu_1^2)(1-\mu^{*2})}},1) ] \notag \\
    &- \sin^{-1} [\min_2 (-1, \frac{\mu_{l,j}-\mu_1 \mu^*}{\sqrt{(1-\mu_1^2)(1-\mu^{*2})}},1) ] \Big) - \frac{d\mu_j}{2} \Big\} \frac{1}{1+\frac{0.1}{p_i}} \Big] \notag \\
    &\times\Big( d\nu_{iz} (w_{r,i}, T_k) - d\nu_{iz} (w_{l, i}, T_k) \Big) F_e(p_k, \mu_{l'}) dp_k d\mu_{l'}- \nu_{iz}^{hard}(p_i) F_e(p_i, \nu_j)
\end{align}
where $d\nu_{iz} = n_H c \tau_c d\sigma_{iz} (w_i, T_k) \beta_k$.
\begin{figure*}
\centering
\begin{subfigure}[b]{0.48\textwidth}
    \centering
    \includegraphics[width=\textwidth]{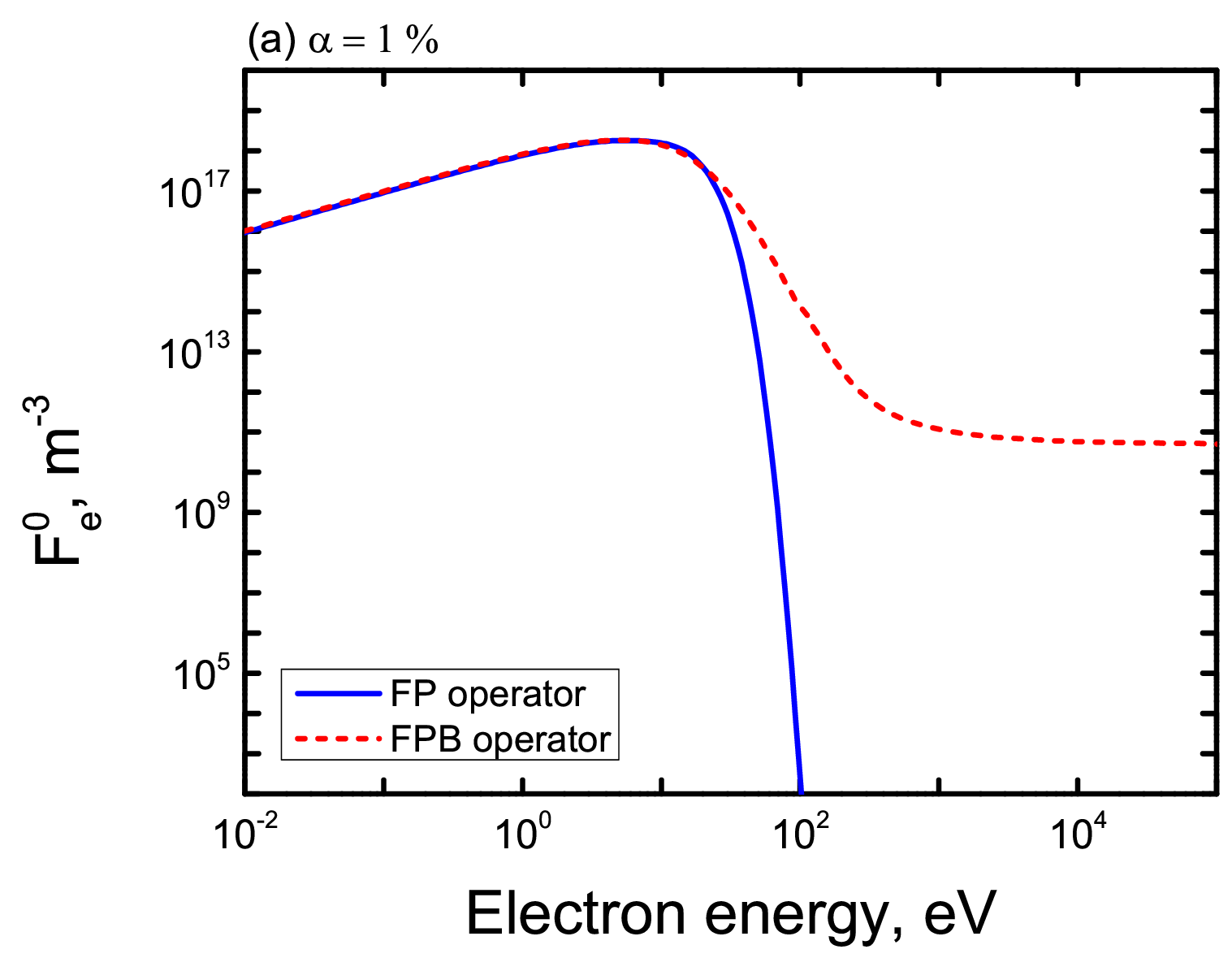}
    \phantomcaption
    \label{fig6a}
\end{subfigure}
\hfill
\begin{subfigure}[b]{0.48\textwidth}
    \centering
    \includegraphics[width=\textwidth]{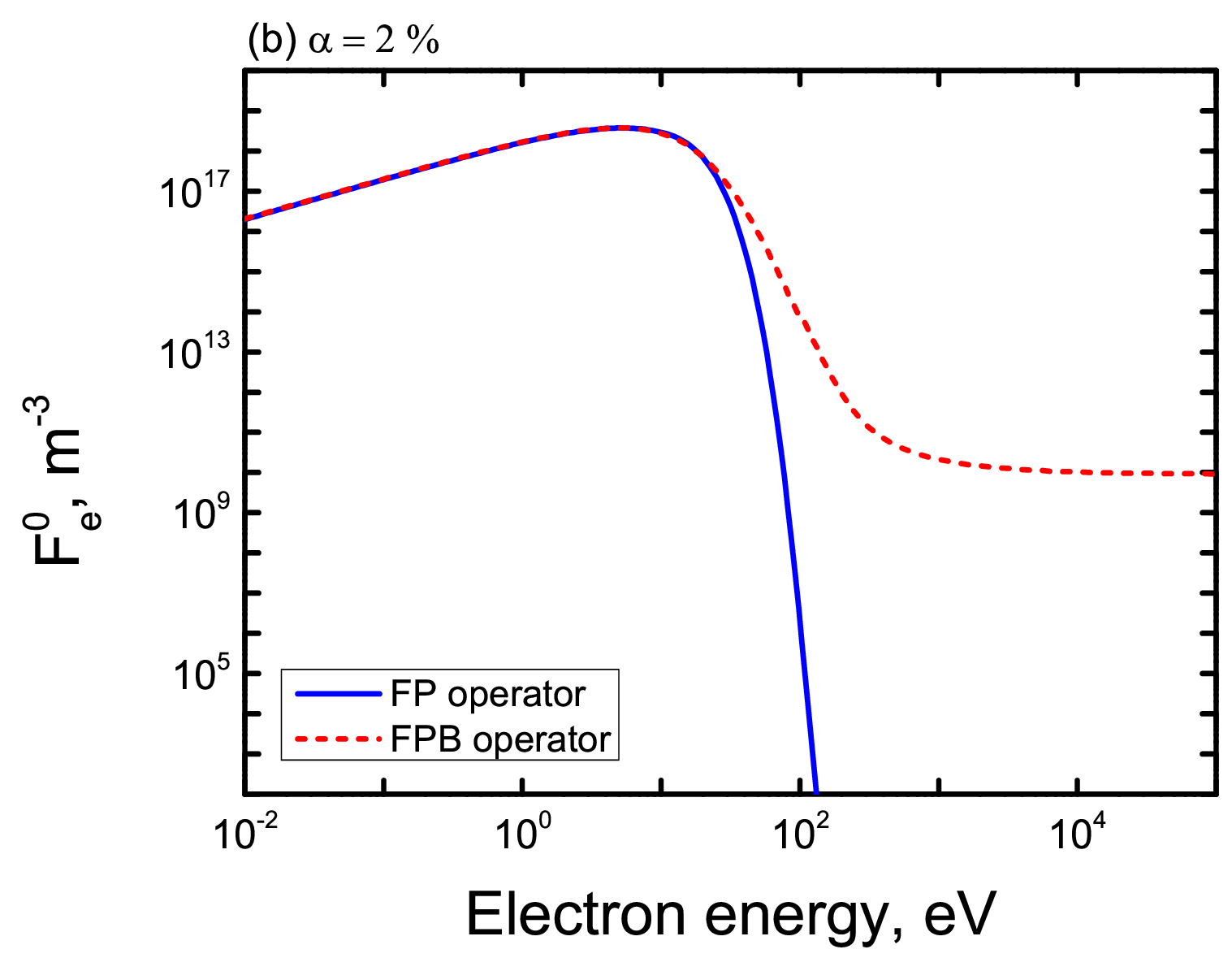}
    \phantomcaption
    \label{fig6b}
\end{subfigure}
\\
\begin{subfigure}[b]{0.48\textwidth}
    \centering
    \includegraphics[width=\textwidth]{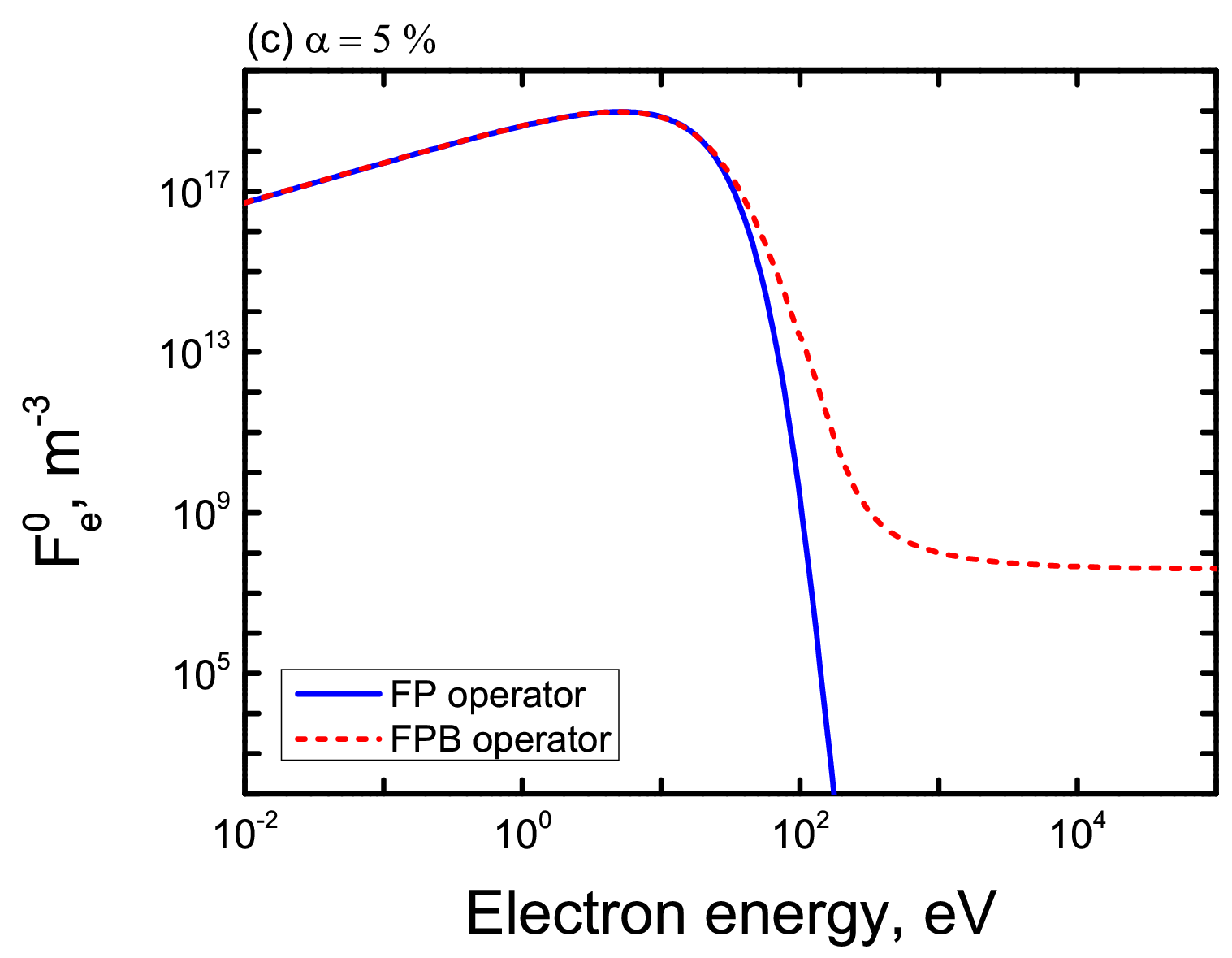}
    \phantomcaption
    \label{fig6c}
\end{subfigure}
\hfill
\begin{subfigure}[b]{0.48\textwidth}
    \centering
    \includegraphics[width=\textwidth]{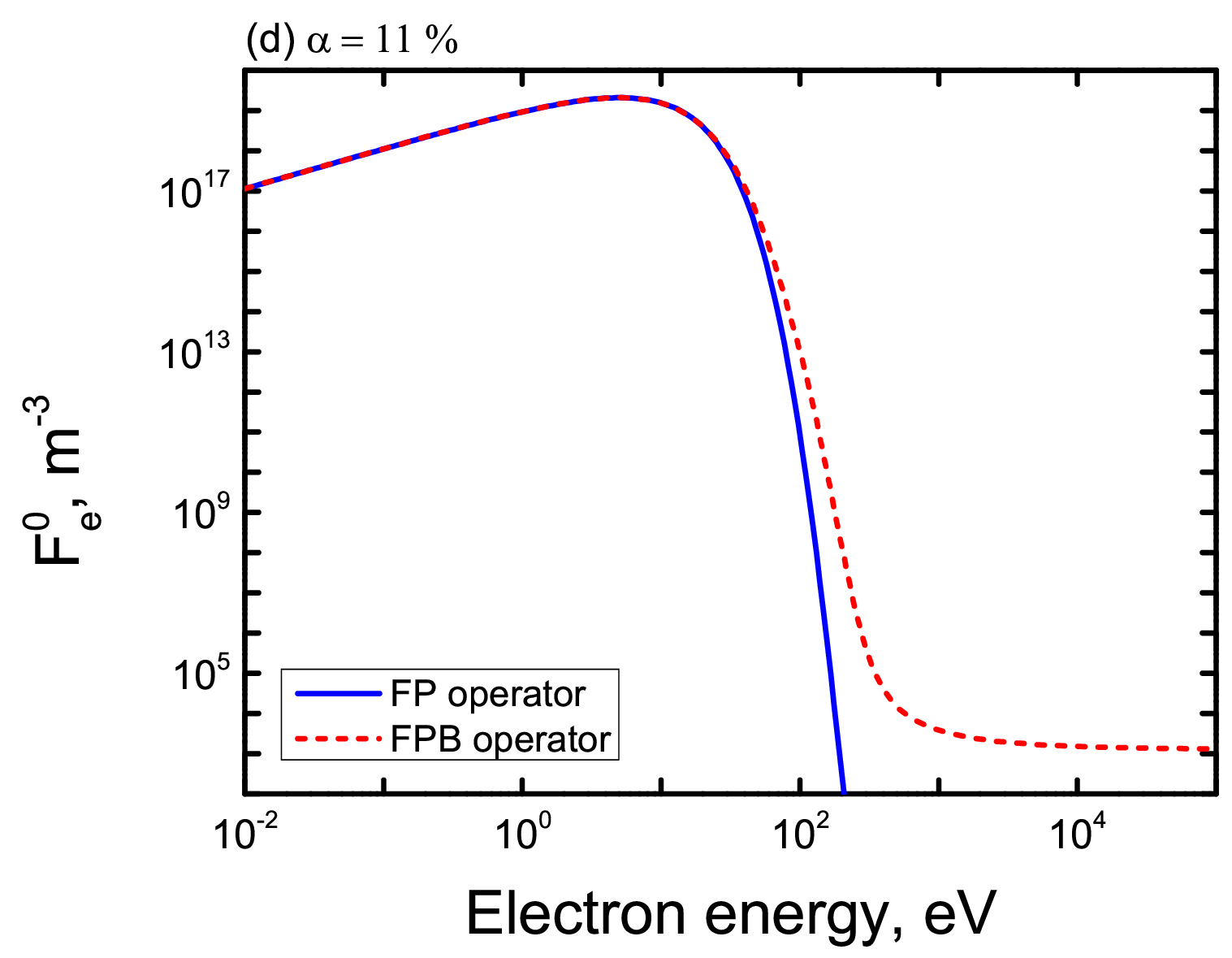}
    \phantomcaption
    \label{fig6d}
\end{subfigure}
\caption{The steady-state solutions of $F_e^0$ with the FP (red dashed curve, h=0.1) and FPB (blue solid curve, h=0.99) operators. $(n_H + n_e) = 10^{18} \ m^{-3}$, $T_e  =  5 \ eV$ and $E=0.2 \ Vm^{-1}$. $\alpha = 1, \ 2, \ 5$ and $11 \ \%$ for (a), (b), (c) and (d), respectively. (a) Reprinted figure with permission from \cite{Lee2024PRL}, copyright 2025 by the American Physical Society.}
\label{fig6}
\end{figure*}
\subsection{Discretization of $\mathcal{B}^{hard}_{ex} \{ F_e \}$}
Consider $\int_{p_{l,i}}^{p_{r,i}} dp$ on the source term of $\mathcal{B}^{hard}_{ex, 1s \to np} \{ F_e \}$ and discretize $p$ integral.

\begin{align}
    \int_{p_{l,i}}^{p_{r,i}} &dp_i \frac{\nu_{ex, 1s \to np}^{hard}(p^+_{1n})}{2} \frac{\beta(p_i)}{\beta(p^+_{1n})} F_e^0(p^+_{1n}) \notag \\
    &= \sum_k \int_{p_{l,i}}^{p_{r,i}} dp_i \int_{p_{l,k}}^{p_{r,k}} dp_k \Big[ \frac{\nu_{ex, 1s \to np}^{hard}(p_k)}{2} \frac{\beta(p_i)}{\beta(p_k)} F_e^0(p_k) \frac{dT_k}{dp_k} \delta (T_k - W_i - E_{1n}) \Big] \notag \\
    &= \sum_k \int_{p_{l,k}}^{p_{r,k}} dp_k \Big[ \frac{\nu_{ex, 1s \to np}^{hard}(p_k)}{2} F_e^0(p_k) \Big] \Big[  \int_{W_{l,i}}^{W_{r,i}} dW_i \delta (T_k - W_i - E_{1n}) \Big]
\end{align}
Here, we use the integral representation as the first step and interchange the order of integration as the second step. Suppose we just take $\int_{W_{l,i}}^{W_{r,i}} dW_i \delta (T_k - W_i - E_{1n}) = 1$. In that case, the particle conservation is adequate but i-th cell-centered energy $W_i$ is not equal to $T_k-E_{1n}$. To overcome this drawback, $\delta (T_k - W_i - E_{1n})$ is smoothed as meeting
\begin{align}
    d\hat{W}^{1n}_{i-1,k} + &d\hat{W}^{1n}_{i,k} + d\hat{W}^{1n}_{i+1,k} = 1, \\
    d\hat{W}^{1n}_{i-1,k} W_{i-1} + d\hat{W}^{1n}_{i,k} W_{i} &+ d\hat{W}^{1n}_{i+1,k} W_{i+1} = T_k + E_{1n},
\end{align}
where $d\hat{W}^{1n}_{i,k} \equiv \int_{W_{l,i}}^{W_{r,i}} dW_i \hat{\delta} (T_k - W_i - E_{1n})$ and $\hat{\delta}$ is the smoothed delta function. If $W_{l,i} < T_k-E_{1n} < W_i$, set $d\hat{W}^{1n}_{i+1,k} = 0$ and then,
\begin{align}
    d\hat{W}^{1n}_{i-1,k} = \frac{W_i - T_k + E_{1n}}{W_i - W_{i-1}}, d\hat{W}^{1n}_{i,k} = \frac{T_k - E_{1n} -W_{i-1}}{W_i - W_{i-1}}.
\end{align}
If $W_{i} < T_k-E_{1n} < W_{r,i}$, set $d\hat{W}^{1n}_{i-1,k} = 0$ and then,
\begin{align}
    d\hat{W}^{1n}_{i,k} = \frac{W_{i+1} - T_k + E_{1n}}{W_{i+1} - W_{i}}, d\hat{W}^{1n}_{i+1,k} = \frac{T_k - E_{1n} -W_{i}}{W_{i+1} - W_{i}}.
\end{align}
If $0 < T_k-E_{1n} < W_{l,0}$, just set $d\hat{W}^{1n}_{0,k} = 1$, assuming the energy loss due to this approximation is negligible. Skipping trivial $\mu$ discretization, take $\sum_i$ to erase $i$ dependence. Then,
\begin{align}
    \sum_{i,j} \int_{p_{l,i}}^{p_{r,i}} &dp_i \int_{\mu_{l,j}}^{\mu_{r,j}} d\mu_j \mathcal{B}^{hard}_{ex, 1s \to np} \{ F_e \} \notag \\
    &= \sum_k \int_{p_{l,k}}^{p_{r,k}} dp_k \Big[ \nu_{ex, 1s \to np}^{hard}(p_k) F_e^0(p_k) \Big] - \sum_i \int_{p_{l,i}}^{p_{r,i}} dp_i \Big[ \nu_{ex, 1s \to np}^{hard}(p_i) F_e^0(p_i) \Big].
\end{align}
We discretize $\int_{p_{l,k}}^{p_{r,k}} dp_k [\nu_{ex,1n \to np}^{hard}(p_k) F_e^0 (p_k)]$ similarly to Eqn. (\ref{eq:disc_pk_izhard}),
\begin{equation}
    \int_{p_{l,k}}^{p_{r,k}} dp_k \nu_{ex, 1s \to np}^{hard}(p_k) F_e^0 (p_k) \approx \nu_{ex, 1s \to np}^{hard}(p_k) F_e^0 (p_k) dp_k. \label{eq:disc_pk_exhard}
\end{equation}
There are two exceptions that we replace: $dp_{k^*}=p_{r,k^*}-p_{l,k^*}$ in Eqns. (\ref{eq:disc_pk_exhard} as $dp_{k^*}=p_{r,k^*}-\sqrt{(E_{1n}'+1)^2-1}$ where $k^*$ satisfies $T_{r,k^*} > E_{1n}$ and $T_{l,k^*} < E_{1n}$ and $dp_{k^{**}}=p_{r,k^{**}}-p_{l,k^{**}}$ as $dp_{k^{**}}=\sqrt{(\frac{E_{1n}'}{h}+1)^2-1}-p_{l,k^{**}}$ where $k^{**}$ satisfies $T_{r,k^{**}} > \frac{E_{1n}}{h}$ and $T_{l,k^{**}} < \frac{E_{1n}}{h}$.
Finally, the fully discretized form of $\mathcal{B}^{hard}_{ex} \{ F_e \}$ is
\begin{align}
    [\mathcal{B}^{hard}_{ex} \{ F_e \}(p_i, \mu_j)]_k &= \sum_n \Big[ \frac{dp_k}{dp_i} \frac{\nu_{ex, 1s \to np}^{hard}}{2} F_e^0(p_k) d\hat{W}_{i,k}^{1n} - \nu_{ex, 1s\to np}^{hard}(p_i) F_e(p_i, \mu_j) \Big]
\end{align}

\begin{figure}
    \centering
    \includegraphics[width=0.8\linewidth]{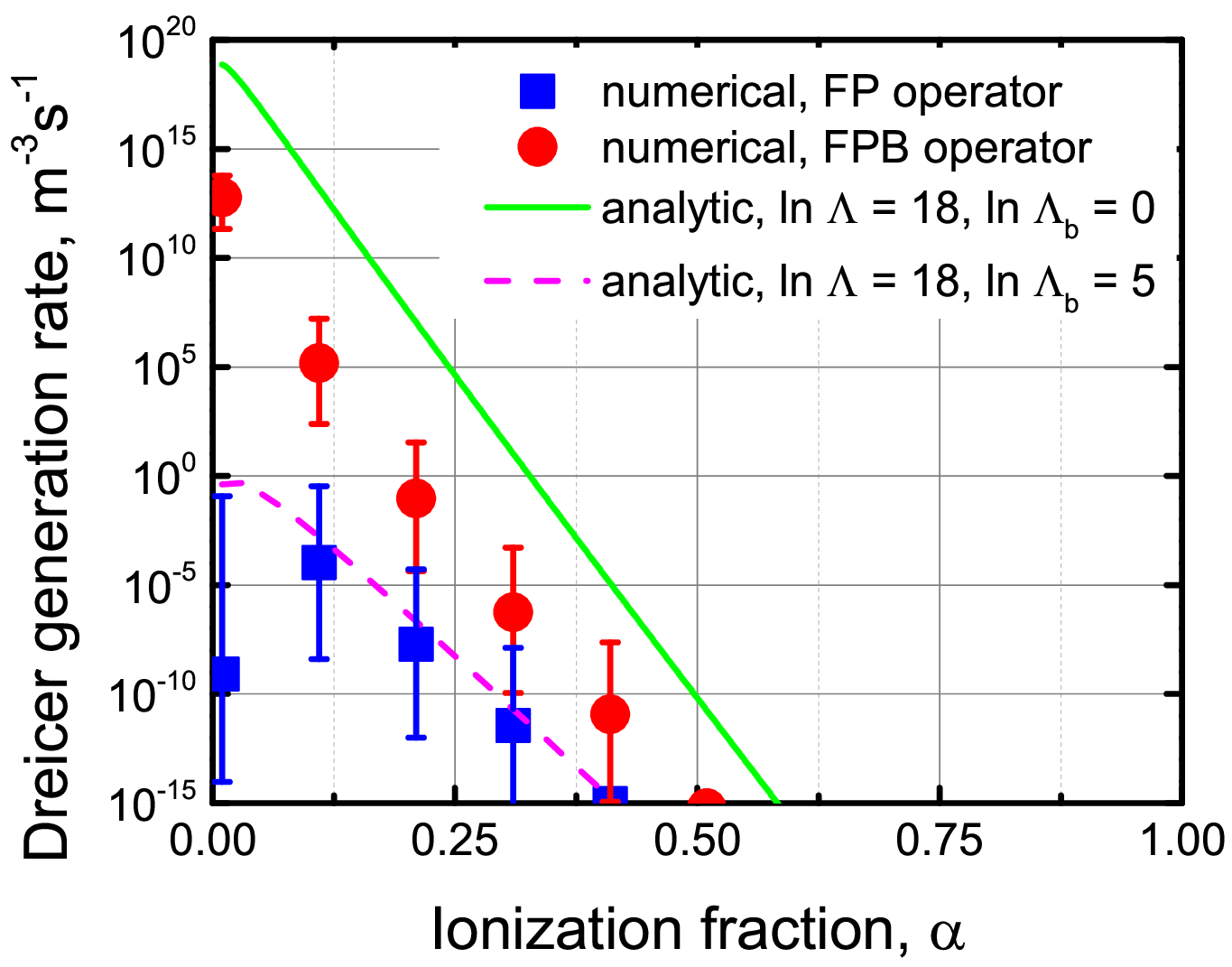}
    \caption{The Dreicer generation rate as a function of $\alpha$: with the FP (blue square, $h=0.99$) and FPB (red circle, $h=0.1$) operators, and the analytic formula \cite{Connor1975NF} with the Coulomb logarithm for free electrons $\ln \Lambda = 18$ and logarithmic factor for bound electrons $\ln \Lambda_{bound} = 0$ (green solid curve) and $\ln \Lambda_{free} = 18$ and $\ln \Lambda_{bound} = 5$ (magenta dashed curve), required for collisional frequency computation. Here, $(n_H + n_e) = 10^{18} \ m^{-3}$, $T_e  =  5 \ eV$ and $E=0.2 \ Vm^{-1}$. Error bars are estimated by scanning $E=0.18 Vm^{-1}$ to $E=0.22$ ($\pm 10 \%$). Reprinted figure with permission from \cite{Lee2024PRL}, copyright 2025 by the American Physical Society.}
    \label{fig7}
\end{figure}

\section{Non-diffusive Dreicer generation in weakly-ionized plasmas \label{sec4}}
In weakly-ionized plasmas, the Dreicer mechanism \citep{Dreicer1959PR} is one of main RE generation mechanisms \citep{Breizman2019NF}. According to this mechanism \citep{Gurevich1961JETP, Connor1975NF}, a diffusive flux forms %the Maxwellian forms 
across the critical momentum $p_c$ under the effect of electric field if electric field $E$ exceeds a critical value $E_c \equiv \frac{e^3 n_e \ln \Lambda_{free}}{4 \pi \varepsilon_0^2 mc^2}$. However, during tokamak startup, $p_c$ can be around several times of the ionization potential or just higher than by one order of magnitude, where inelastic collisions are predominantly hard. In this case, the binary nature of inelastic collisions facilitate that some electrons experience virtually no inelastic collisions during acceleration to $p_c$. The binary nature can be fully accounted by the Boltzmann operator, Eqn. (\ref{eq:hard}) by selecting an appropriate value of $h$ as the ratio of the characteristic excitation energy to the critical runaway energy of interest, i.e. $\approx 0.1$ \citep{Lee2024PRL}.

The non-diffusive electron acceleration mechanism that was discussed in Sec. \ref{sec2-5} is demonstrated in Fig. \ref{fig6}, where the steady-state solutions are illustrated for various ionization fractions (corresponding to Fig. \ref{fig4}). For the demonstration, we only took the test particle part to eliminate RE avalanche effect \citep{Rosenbluth1997NF} by setting upper boundary of integral in Eqn. (\ref{eq:hard}) to $\sqrt{(2w+b'+2)(2w+b')}$. Blue curves in Figs. \ref{fig6a}-\ref{fig6d} represent the solutions where most collisions are described with the FP operator (h=0.99). In all cases, the resulting $F_e^0 < 10^{3} m^{-3}$ with electron energy above $1 \ keV$. Indeed, no Dreicer flow is expected in such conditions according to effectiveness of Dreicer generation \citep{Jayakumar1993PRA} $E/E_d^{eff} < 0.02$ with $E_d^{eff} \equiv \frac{e^3 (n_e \ln \Lambda + n_H \ln \Lambda_b)}{4 \pi \varepsilon_0^2 T_e}$ where an effect of free-bound collisions are considered by replacing $n_e \ln \Lambda_{free}$ as $n_e \ln \Lambda_{free} + n_H \ln \Lambda_{bound}$ \citep{Martinsolis2015PoP} and $\ln\Lambda_{bound} \approx \ln\Lambda^{tot} \approx 5$ (see Fig. \ref{fig2}). The red dashed curve shows the solution with the FPB collision operator with $h=0.1$. This solution is close to Maxwellian in lower energy where free-free collisions dominate. But it differs significantly at higher energy where binary Boltzmann collisions allow for virtually free acceleration for some particles. These freely accelerated particles become runaways. The more $\alpha$ is the less REs are created since continuous free-free collisions induce stronger frictions than free-bound collisions.

Figure \ref{fig7} shows Dreicer generation rate as a function of the ionization fraction. The corresponding parameters are close to that of standard Ohmic discharge in KSTAR tokamak \citep{Yoo2018Nat,Lee2023PS} and ITER plasma initiation \citep{Vries2019NF}. In this case, the binary nature of collisions plays a critical role in RE generation as indicated by the condition $E/E_d > 0.025$ but $E/E_d^{eff} < 0.02$: the conventional Dreicer generation is ineffective for these parameters while the Dreicer generation accounted by Boltzmann operator is effective and results in $10^{13} \ m^{-3} s^{-1}$ for $\alpha = 0.01$. This is consistent with the analytical indicator of effectiveness since $E/E_d \approx 0.43 \gg 0.02$ but $E/E_d^{eff} \approx 0.015$. The observed rates are much lower than the analytic prediction by Connor-Hastie \citep{Connor1975NF} without free-bound collisions (green curve in Fig. \ref{fig7}). Our calculations for the conventional Dreicer generation (blue squares) agree with the Connor-Hastie formula if $E_d^{eff}$ with $\ln \Lambda_{b} \approx 5$ deduced from $\ln \Lambda^{tot}$ is used in the formula (dashed curve).

\section{Conclusion}
The collision operator of electrons in cold weakly-ionized plasmas is developed to analyze electron runaway process. The FPB operator for free-bound collisions is derived with the help of valid collision cross sections for elastic \citep{Itikawa1974ADNDT, Buckman2000ECA, Jablonski2004JPCRD, Salvat2005CPC, Breizman2019NF} and inelastic collisions \citep{Kim1994PRA, Kim2001PRA, Kim2000PRA}. The operator is invariant in particle and energy loss of electrons with respect to $h$ since the integral boundary to define the logarithmic factor is appropriately selected \citep{Embreus2018JPP}, which in turn ensures the particle and energy conservation. For free-free collisions, validity of the linearized collision operator with sufficiently low ionization fraction ($\sim$ middle phase of breakdown \citep{Yoo2017CPC, Yoo2018Nat, Yoo2022PPCF} or early burn-through \citep{Kim2012NF}) is verified by elucidating that the Maxwell distribution function can be sustained under a strong effect of neutrals.

For numerical simulation, we present the complete analytic expression of the volume averaged form of the FPB operator. This can be implemented in the FVM-based numerical solver such as FiPy \citep{Guyer2009CSE}.

Employing the FPB operator, we generalize the Dreicer generation model for weakly-ionized plasmas to the non-diffusive regimes. An essential role of the binary nature of inelastic collisions is played in the Dreicer generation. Therefore, this work is envisaged to provide the precise RE generation rate for designing of a runaway-free reactor startup.

One of promising applications of this work is to couple the kinetic model to "fluid" plasma burn-through simulation \citep{Kim2012NF, Kim2022NF} like fluid-kinetic disruption RE simulation \citep{Aleynikov2017NF, Hoppe2021CPC}. This coupling will facilitate self-consistent simulation of runaway current evolution as well as fully kinetic description of ionizing avalanche. Moreover, the electron cyclotron pre-ionization \citep{Lloyd1991NF} will be investigated in a kinetic level by coupling this with nonlinear electron-cyclotron energy gain model \citep{Suvorov1988SJP, Farina1991PoFa, Farina1991PoFb, Seol2009PoP, Farina2018NF, Johansson2024JPP, Gwak2025NF}.

\section*{Acknowledgement}
Yeongsun Lee is grateful for fruitful discussion to Dr. Hyun-Tae Kim, Dr. Jeongwon Lee, Prof. Mathias Hoppe, Mr. Jinwoo Gwak and Prof. Gyungjin Choi. The views and opinions expressed herein do not necessarily reflect those of the ITER Organization. This research was supported by National R\&D Program through the National Research Foundation of Korea (NRF) funded by Ministry of Science and ICT (2021M3F7A1084419).

% susie put cite commands here, don't bother with citet etc just yet.

\bibliographystyle{jpp}

\bibliography{ref}
%\section*{Reference}
%\bibliographystyle{unsrt}
%\bibliography{ref}

\end{document}